\documentclass[preprints,article,accept,oneauthor,pdftex]{mdpi} 
\firstpage{1} 
\pubvolume{xx}
\issuenum{1}
\articlenumber{5}
\pubyear{2020}
\copyrightyear{2020}
\history{Received: date; Accepted: date; Published: date}

\def\XXint#1#2#3{{\setbox0=\hbox{$#1{#2#3}{\int}$}
     \vcenter{\hbox{$#2#3$}}\kern-.5\wd0}}

\newcommand{\bmu}{\bar{\mu}}

\newcommand{\ep}{\epsilon}
\newcommand{\vep}{\varepsilon}
\newcommand{\ve}{\varepsilon}
\newcommand{\vh}{\varphi}

\newcommand{\vp}{{\mathbf{p}}}

\newcommand{\vq}{{\mathbf{q}}}

\newcommand{\cT}{{\cal T}}

\newcommand{\la}{\langle}
\newcommand{\ra}{\rangle}

\newcommand{\nn}{\nonumber}

\newcommand{\zh}{{\rm th}}

\Title{Unitarization Technics in Hadron Physics with Historical Remarks}

\Author{Jos\'e Antonio Oller\orcidA{}}

\AuthorNames{J. A. Oller}

\address{
Departamento de F\'{\i}sica. Universidad de Murcia. E-30071,
Murcia. Spain; oller@um.es}

\corres{Correspondence: oller@um.es}

\abstract{We review a series of unitarization techniques that have been used during the last decades, 
  many of them  in connection with the advent and development of current algebra and later of
  Chiral Perturbation Theory. 
  Several methods are discussed like the generalized effective-range expansion,
  $K$-matrix approach, Inverse Amplitude Method, Pad\'e approximants and the $N/D$ method.
  More details are given for the latter though. We also consider how to implement them in
  order to correct by final-state interactions. In connection with this some other methods are
  also introduced like
  the expansion of the inverse of the form factor, the Omn\' es solution, generalization to coupled
  channels and the Khuri-Treiman formalism, among others.}

\keyword{Unitarity; Scattering; Partial-wave amplitudes; Final-state interactions; Analyticity; $S$-matrix theory; Spectroscopy; $N/D$ method; Left-hand cut.}

\begin{document}


\tableofcontents

\newpage

\section{Introduction}
\label{sec:intro}
\def\theequation{\arabic{section}.\arabic{equation}}
\setcounter{equation}{0}

The effective chiral Lagrangian formalism has become a well-established methodology to study the interactions 
of the Goldstone bosons  with or without other particle species, like e.g. pions and nucleons, respectively
\cite{ccsc}. 
The most significant example is Chiral Perturbation Theory (ChPT) \cite{weinberg.200221.1,leutwyler.200220.1},
which is the low-energy effective field theory (EFT) of Quantum Chromodynamics (QCD).
For introduction and reviews, see e.g. \cite{u1,u2,u3}. 
The use of perturbative calculations within ChPT as input for
non-perturbative $S$-matrix based methods is a general procedure several decades old.  
Due to the fact that ChPT results are perturbative, given in terms of an expansion 
organized in increasing powers of the external four-momenta and light pseudoscalar masses, 
unitarity is only satisfied in the perturbative sense, similarly as in a standard Born series
(perturbative unitarity is discussed in Sec.~\ref{sec.200525.1}).
A well-known  example in this regard is the calculations in Quantum Electrodynamics with Feynman diagrams,
where the expansion is done in powers of $\alpha$ (the fine structure constant),
so that if the leading-order calculation is ${\cal O}(\alpha)$ then unitarity contributions
start at ${\cal O}(\alpha^2)$ 
from one-loop diagrams.
However, the fulfillment of unitarity implies to square the calculated amplitudes,
and not to expand the latter only up to the order in which the scattering amplitude is calculated
(for an explicit example the reader can consult e.g. Sec.~7.3 of Ref.~\cite{peskin}.) 

It is somewhat astonishing that already in 1970 one can read about motivations 
for unitarizing phenomenological chiral Lagrangians, introduced
to construct realizations of the current algebra approach \cite{weinberg.200221.1}.
Rephrasing  the original remarks by Schnitzer \cite{schnitzer.200221.1,schnitzer.200302.1}, the ideas he put forward are
still the main reasons to advocate the unitarization of ChPT amplitudes:

\begin{enumerate}
\item The tree approximation to the scattering amplitudes violate badly unitarity. This could also be  
said for perturbative unitarity, at least in some partial waves.
\item The Lagrangians are nonlinear and nonrenormalizable, which makes difficult to compute higher-order corrections.  
Nowadays, we would better say that there is a rapid proliferation of 
counterterms as the order of the calculation increases in ChPT, with the state of the art at the
two-loop level in ChPT. 
It is typically simpler and much more predictive to implement lower-order 
calculations of ChPT within  non-perturbative methods.\footnote{
Several examples  are given along this review
related to meson-meson scattering and spectroscopy, like e.g. the $\pi\pi$ phase shifts, scalar and vector
pion form factors, impact of the resonances $f_0(500)$ and $\rho(770)$ in the low-energy phenomenology,
$\eta\to 3\pi$ decays, etc.
}

\item Even if such corrections could be computed, the resultant renormalized perturbation series would 
probably diverge, since the perturbation parameter has the strength characteristic of strong interactions. 
This is clear from phenomenology because hadronic interactions are characterized by 
plenty of resonances and a rapid saturation of unitarity in many partial-wave amplitudes (PWAs). 

\end{enumerate}

Although the interest in the present writing is on ChPT and the associated chiral expansion, among those
early papers of Schnitzer we also quote Ref.~\cite{schnitzer.200302.1}. 
This paper builds  a particular realization of the current-algebra, which
satisfies the associated Ward-Takahashi identities
and two-body unitarity is implemented by means of an effective-range-type parameterization (a
unitarization method discussed in Sec.~\ref{sec.200523.1}).

 One possibility to improve the agreement with data of the perturbative calculations within ChPT is
to apply the chiral series expansion to an interaction element of the amplitude, which
is afterwards implemented within non-perturbative techniques.
This is one of the basic ideas behind unitarization methods for the chiral series of
scattering amplitudes. 
The first works along these lines considered the application of an
effective-range-type parameterization
to unitarize $\pi\pi$ scattering \cite{brown.200302.1,schnitzer.200302.1},
once the $\pi\pi$ scattering amplitude was calculated at leading order in ChPT
by the application of the current-algebra techniques and
the partial conservation of the axial-vector current (PCAC) \cite{weinberg.200302.1,brown.200302.1}.
A similar unitarization method was later applied to the first calculation at next-to-leading order (NLO) 
in the chiral counting of the $\pi\pi$ PWAs in the chiral limit ($m_\pi=0$).
The calculation of the latter ones, as well as their
unitarization by applying a generalized effective-range expansion (ERE) \cite{chew},
were undertaken in Ref.~\cite{lehmann.200302.1},
as discussed in more detail in Sec.~\ref{sec.200511.1}.
This calculation explicitly shows that the $\pi\pi$ scattering in the chiral limit is finite.


The pioneering works by Truong and collaborators \cite{truong.200211.1,truong.200211.2,truong.200211.3}
deserve special mention,
in which the role of the isoscalar $S$-wave $\pi\pi$ final-state interactions (FSI) are stressed,
having significant effects on several physical processes.
In a first instance  \cite{truong.200211.1}, the authors correct the current-algebra result for $\eta\to 3\pi$
by the $S$-wave $\pi\pi$ rescattering, a reaction which is also discussed by the application of the
Khuri-Treiman (KT) \cite{khuri}
formalism in Sec.~\ref{sec.200522.1}.
For that Ref.~\cite{truong.200211.1} multiplies the current-algebra transition
amplitude by an Omn\`es function \cite{oller.book,musk.book},
in which the isoscalar scalar $\pi\pi$ phase shifts were, however, taken from experiment.
As a result, the Watson final-state theorem is fulfilled \cite{watson.200211}. 
In another work of the saga \cite{truong.200211.2}, the input phase shifts were generated consistently by the
theoretical scheme followed after taking the one-loop ChPT result for the scalar and vector pion form factors and
imposing the fulfillment of unitarity, as discussed in Sec.~\ref{sec.200515.1}.
It is also stressed that in this form a resummation of the ChPT series is achieved
that may also give rise to resonant effects.


It is also remarkable the confirmation by unitarization methods
of the existence of the $\sigma$ resonance in pion-pion interactions
at low energies. This resonance is nowadays called $f_0(500)$ in the PDG \cite{pdg} and its pole position is
given there at $(400-550)-i(200-350)$~MeV. The standard view of ChPT,
based on the spontaneous symmetry breaking of $SU(2)\times SU(2)$ chiral symmetry \cite{leutwyler.200220.1},
considered as highly unlikely that such a low-mass resonance could happen in $\pi\pi$ scattering, 
where the small expansion parameter is claimed to be $m_\pi^2/m_\rho^2\simeq 0.03$, with  $m_\rho$ the mass of the
$\rho(770)$ meson. However, for the isoscalar scalar $\pi\pi$ scattering the unitarity corrections are affected by a 
large numerical factor that could actually make  the  expansion parameter in the momentum-squared dependence of
these PWAs to be much larger.
This was explicitly shown in Ref.~\cite{nd} by performing the exercise of determining the value of the
renormalization scale $\mu$ needed in order to generate the $\rho(770)$ pole by unitarizing the leading-order (LO) 
ChPT amplitude. It was obtained that a huge unnatural value for the realm of QCD was needed, with $\mu \simeq 1$~TeV,
while the same value for generating the $\sigma$ resonance had the natural value in QCD of $\mu\simeq 1$~GeV.

It is instructive to also show the main equations for the completion of this exercise. For the isovector vector 
$\pi\pi$ interactions, where the $\rho(770)$ resonance appears, one has the unitarized expression of the
LO ChPT PWA $T_{11}(s)$, which reads \cite{nd} (as also discussed in Sec.~\ref{sec.200517.1}) 
\begin{align}
\label{200508.1}
T_{11}(s)&=\left[\frac{6f_\pi^2}{s-4m_\pi^2}+g(s)\right]^{-1}~.
\end{align}
Here, $T_{IJ}$ is the PWA of the two-pion system with isospin $I$, $J$ is the angular momentum and 
the LO ChPT amplitude is $(s-4m_\pi^2)/6f_\pi^2$, with $f_\pi=92.4$~MeV the pion weak decay constant. 
The function $g(s)$ corresponds to the two-pion unitarity loop function, given by
\begin{align}
\label{200508.2}
g(s)&=\frac{1}{(4\pi)^2}\left[\log\frac{m_\pi^2}{\mu^2}+\sigma(s)\log\frac{\sigma(s)+1}{\sigma(s)-1}\right]~,\\
\sigma(s)&=\sqrt{1-4m_\pi^2/s}~.\nn
\end{align}
In turn, the unitarized expression for the $I=J=0$ PWA, which contains the $f_0(500)$ pole, is \cite{npa,nd}
\begin{align}
\label{200509.3}
T_{00}(s)&=\left[\frac{f_\pi^2}{s-m_\pi^2/2}+g(s)\right]^{-1}~,
\end{align}
where the LO ChPT PWA is $(s-m_\pi^2/2)/f_\pi^2$.  
The main difference between Eqs.~\eqref{200508.1} and \eqref{200509.3} is the factor 6 dividing the LO ChPT $T_{11}(s)$ compared to
$T_{00}(s)$, because $s\gg m_\pi^2$ in the region where the $\sigma$ or $\rho$ poles lie. 
Indeed, in order to get a resonance of mass $m_\rho$ in the  $I=J=0$ PWA one needs a $\mu$ of around 1.8~GeV,
in comparison with $\mu$ around 1~TeV that is needed in the $I=J=1$ case.
The reason for this dramatic change in the needed values of $\mu$ is because  $g(s)$ only depends logarithmically on
this parameter. 
This fact reflects that the unitarity corrections for the scalar isoscalar sector are numerically enhanced. 
This 
enhancement is enough to generate resonant effects that strongly impact the phenomenology 
and make fallacious to think in the possibility to reach accuracy by a straightforward application
of ChPT for many reactions. As a result, the infinite set of unitarity bubble diagrams should be resummed 
in order to account for this numerical enhancement.

This phenomenon is also seen in the strong corrections affecting the $I=J=0$ $\pi\pi$
scattering length originally calculated by Weinberg at LO \cite{weinberg.200302.1} with current algebra methods. 
The expressions for the $a_{IJ}$ scattering lengths up to NLO or ${\cal O}(p^4)$ in ChPT 
from Ref.~\cite{leutwyler.200220.1} contain chiral loops which are the dominant NLO contributions in the limit
$M_\pi\to 0$. They read:
\begin{align}
\label{200509.1}
a_{00}&=\frac{7m_\pi^2}{32\pi f_\pi^2}\left\{1-\frac{9m_\pi^2}{32\pi^2 f_\pi^2}\log\frac{m_\pi^2}{\mu^2}+\ldots\right\}~,\\
a_{20}&=-\frac{m_\pi^2}{16\pi f_\pi^2}\left\{1-\frac{5m_\pi^2}{32\pi^2 f_\pi^2}\log\frac{m_\pi^2}{\mu^2}\ldots\right\}~,\nn\\
a_{11}&=\frac{1}{24\pi f_\pi^2}\left\{1-\frac{m_\pi^2}{8\pi^2f_\pi^2}\log\frac{m_\pi^2}{\mu^2}+\ldots\right\}~.\nn
\end{align}
It follows then that $a_{00}$ has the largest NLO contribution in the limit $m_\pi\to 0$. In order to
appreciate better the relatively large size of this correction, it is worth comparing it with 
the pion mass calculated in ChPT  up to NLO \cite{leutwyler.200220.1},
\begin{align}
\label{200509.2}
m_\pi^2&=\bar{m}_\pi^2\left\{1+\frac{m_\pi^2}{32\pi^2 f_\pi^2}\log\frac{m_\pi^2}{\mu^2}+\ldots\right\}~,
\end{align}
with $\bar{m}_\pi^2$ the bare mass squared.
The NLO term here is a factor 9 smaller in absolute value than that for $a_{00}$. 
Indeed, this was one of the reasons for developing a non-perturbative dispersive approach
that could provide an improvement in the prediction of the $\pi\pi$ scattering lengths.
The idea is to make use of the Roy equations \cite{roy.200509.1} and to match with ChPT
in the subthreshold region, where the ChPT expansion is better behaved, 
since it is away from the threshold cusps \cite{anant,leutwyler.200509.1}. In this way, the two subtraction constants
needed for solving the Roy equations can be predicted by ChPT, applied at different orders. The resulting
convergence properties of the prediction for the $\pi\pi$ scattering lengths is much improved and a
reliable estimate of the uncertainties can be also provided.
Another more recent advance was the development of a new set of Roy-like equations
in Refs.~\cite{217.pela,111.pela}, the so-called GKPY equations.
The difference is that these equations have only one subtraction instead of two and, e.g.,
they have given rise to an accurate determination  of the $f_0(500)$ pole
from experimental data, without relaying on the ChPT expansion.

These unitarity techniques have also  other interesting fields of application beyond meson physics.
 Indeed, 
the 90's of the past century experienced a boost in the interest of applying chiral 
EFTs for the study of nuclear interactions.
To large extent this was triggered by the seminal articles of Weinberg \cite{weinberg.200301.1},
in which  the systematic application of ChPT order by order to calculate the nuclear potentials $V$ is established. 
As the chiral order increases, however, extra derivatives with respect to $r$ act on the potential, so that
it becomes more singular for $r\to 0$. 
Because of this complication the application of ChPT for the calculation of 
the low-energy $NN$ PWAs  by implementing the chiral potentials in quantum-scattering integral equations 
is not yet fully satisfactory.
In atomic and molecular physics the scattering by a singular potential is of great importance too, a well-known
example being the Van der Waals force among atoms or molecules.  
This is a problem in which recent advances giving rise to the exact $N/D$ method in non-relativistic scattering
\cite{oller.lhc.aop,oller.lhc.plb}  
are showing themselves as very powerful and  promising.
This new method is briefly reviewed in Sec.~\ref{sec.200517.4}. 
The application of ChPT with barons to  $NN$ scattering also triggered the use of this EFT to
the study of the non-perturbative $\bar{K}N$ scattering in coupled channels, particularly in connection with the
$\Lambda(1405)$ \cite{kaiserweise,or,plb,ooj}.

The non-perturbative character of the $ NN $ interactions, which requires the full iteration of the potential, 
is due to two basic aspects.
(i) One of them is a quantum effect of kinematical origin within the typical scales of the problem.
The typical distance of propagation of two nucleons as virtual particles is 
$l_{NN} \sim 1/ E_{\text{kinetic}} \sim m/p^2 \sim (m/m_\pi) b_\pi $, where $m$ and $m_\pi$ are the
nucleon and pion masses, respectively.
The range of the $ NN $ interactions is given by the  Compton wavelength of the pion
$b_\pi=m_\pi^{-1}$ (in our units $\hbar=c=1$). 
As $ m/m_\pi\gg 1 $ this travel distance for virtual particles
is large enough for having several repetitive collisions  
between the propagating two nucleons. 
The same conclusion is reached if one focuses on the propagation of real nucleons.
For a typical  three-momentum $m_\pi$ they have a velocity of order $m_\pi/m$. 
Thus, the time for crossing a distance $b_\pi$ is $\sim m/ m_\pi^2\gg b_\pi$. 
(ii) Nonetheless, if the coupling between two nucleons  
were small enough the scattering  would be perturbative despite (i). 
This does not happen since the coupling due to one-pion exchange  between two nucleons 
is of the order $g_A^2 m_\pi^3/f_\pi^2$, where 
$g_A\simeq 1.26$ is the axial coupling of the nucleon. 
This factor times $l_{NN}$ implies the dimensionless number
\begin{align}
\label{180507.2}
\frac{l_{NN}}{16\pi}\frac{g_A^2 m_\pi^3}{f_\pi^2}=\frac{m g_A^2}{16\pi f_\pi^2} m_\pi~,
\end{align}
which is about  0.5. Therefore, the $NN$ interactions should be treated non-perturbatively as a general rule.  
In this equation the phase-space factor $1/16\pi$ is included,
which accounts for the two-nucleon propagation in all directions.
We also distinguish in Eq.~\eqref{180507.2} the scale \cite{oller.180722.1,birse.180506.1}
\begin{align}
\label{180613.1}
\Lambda_{NN}= \frac{16\pi f_\pi^2}{m g_A^2}\simeq 2 m_\pi~,
\end{align}
which has a striking  small size despite it is not proportional to $m_\pi$.
This is another consequence of the non-perturbative character of the $NN$ interactions.

The unnaturally large size of the $NN$ $S$-wave scattering lengths ($a_s$), 
so that they are much bigger in absolute value than the Compton wavelength of the pion, $|a_s|\gg b_\pi $, 
introduces a new scale at low energies. 
For instance, the scattering length for the isovector $^1S_0$ $NN$ PWA is $a_s \simeq -25$~fm. 
As a result,  the dimensionless number in Eq.~\eqref{180507.2}  becomes even larger  
by a factor $|a_s|m_\pi \gg 1 $.
Therefore, when the center of mass (CM) three-momentum is smaller than  $m_\pi$,
in which case the ERE applies, the $NN$ interactions are manifestly non-perturbative
and the $NN$ potential has to be iterated. Precisely, in this energy region one finds the
bound state of the Deuteron in the coupled ${^3S}_1-{^3D}_1$ PWAs and an antibound state for the $^1S_0$.


One close field is infinite nuclear matter,
where resummation techniques based on the $N/D$ method, discussed in Sec.~\ref{sec.200517.1},
were applied in Refs.~\cite{lacour,llanes} to work out the $NN$ scattering amplitude
in the nuclear medium. From this result, 
equations of state for neutron and symmetric nuclear matter were derived each containing only a 
free parameter, and showing themselves as very successful from the phenomenological point of view.
See Ref.~\cite{oller.nm.review} for a recent review on these and other connected works.
Related resummations were achieved in Refs.~\cite{kaiser,boulet} to address the unitary limit in normal
nuclear matter for a Fermi Gas.
This issue concerns both nuclear physics, condensed matter and atomic, molecular, and optical physics.

At higher energies, one also finds examples of the application of unitarization techniques, some of them, like the
$N/D$ method or the Inverse-Amplitude Method (IAM), discussed here.
Regarding this point, there have been recently a series of works applying these two methods to study the
scattering and spectrum of the longitudinal components of the electroweak gauge vector bosons $W$ and $Z$
by taking advantage of the equivalence theorem, which is applicable to energies much larger than the masses
of the $W$ and $Z$ bosons \cite{f1,f2,f3,f4}. These studies are very timely due to the experimental program at the
LHC, which reinforces their interest.

Quantum gravity is another field in which unitarization techniques have been applied in the last years to
study the $2\to 2$ scattering, due to one-graviton exchange in the $s$ channel,  of   $N_S$ scalars, $N_V$ vector and
$N_f$ fermions. 
Notice that this set of fields comprises all the particles in the standard model as a particular case.
The two particles making up an initial or final two-body state are selected  so as to avoid the graviton $t$- and/or $u$-channel
exchanges. The reason is because these exchanges drive to infrared divergences (gravity is a force of infinite range)
that invalidate a standard
partial-wave amplitude expansion. The quantum corrections are implementing within the low-energy EFT of 
Quantum Gravity \cite{weinphys,burgess,donoghue.200205.1}.
The interested reader can consult Refs.~\cite{han.200204.1,donoghue.200204.1,calmet.200204.1}. 
These works employ the one-loop vacuum polarization due to matter fields (gravitons are excluded),
and resum its iteration plus the tree-level contribution. 
 Of course, a similar situation also
arises in the electromagnetic case by the exchange of a photon in the $t$- and $u$-channels.
A prominent example of it being the Coulomb scattering. 
An interesting future prospect is to develop unitarization methods appropriate for infinite-range interactions.
It could then handle crossed one-graviton exchanges and allow to study those  $2\to 2$  scattering processes
disregarded in Refs.~\cite{han.200204.1,donoghue.200204.1,calmet.200204.1}.

In this work we review a  set of unitarization methods and we always
follow the order of first discussing scattering, mostly in PWAs, and then FSI.
We also develop links between the different methods discussed.
The unitarization techniques selected are popular ones within the hadron physics community.
One of the reasons for their popularity is because they have proven to be very powerful
in phenomenological applications, so that they are certainly of interest.
It was not the aim of this work to be exhaustive and give a comprehensive review
discussing every unitarization method used in the literature.
Historical reasons are behind the inclusion of the (generalized) ERE unitarization, widely used in the  
earlier papers of the 60's and 70's, since later on this method  was replaced by the IAM, 
$K$-matrix parameterizations, $N/D$ method, etc,  in relativistic hadron-hadron scattering (not so for non-relativistic applications).

The contents of this work are organized as follows.
After a brief review on the $S$-matrix and unitarity in Sec.~\ref{sec.200525.1} we then move to discuss several
unitarization methods in Sec.~\ref{sec.200511.1}. 
The generalized ERE, the $K$-matrix approach, the IAM
and the Pad\'e resummation are then considered. The Sec.~\ref{sec.200523.4} is dedicated to
the implementation of re-scattering effects in probes and several methods are presented, with some of them clearly
related to the already presented ones in Sec.~\ref{sec.200511.1} 
dedicated to scattering. Subsequently, other methods are introduced that could be applied to any given set of PWAs.
We discuss in Sec.~\ref{sec.200517.1} the $N/D$ method for PWAs and FSI. This section ends with a brief
account of the exact $N/D$ method recently developed for non-relativistic scattering. 
The last section contains our conclusions with extra discussions included.

\section{Unitarity}
\setcounter{equation}{0}
\label{sec.200525.1}

The $S$-matrix operator $S$ gathers the transition probability amplitudes between in and out states
in a scattering process. 
Let us denote  by $|\alpha\ra_{\rm in}$  and 
$|\beta\ra_{\rm out}$ an `in'  and an `out' state  in the Heisenberg picture,
respectively.
Then, the matrix elements of the $S$ matrix, $S_{\beta\alpha}$, correspond to the scalar products  
\begin{align}
\label{200608.1}
S_{\beta\alpha}&={_{\rm out}\la}\beta|\alpha{\ra_{\rm in}}~.
\end{align}

The $S$ matrix plays  a central role in Quantum Field Theory (QFT) \cite{weinberg.vol1}. 
One is typically concerned with the matrix elements of the $S$-matrix so as
to extract scattering observables out of a QFT. 
A crucial property in this regard is that the (on-shell) matrix elements of the $S$-matrix are   
invariant under reparameterization of the quantum fields in QFT \cite{haag.58,ccsc}.

The analytical continuation of the $S$ matrix in the complex-energy plane allows to determine the
spectrum of the theory. Its continuum part corresponds to branch cuts and the
bound states, virtual states and resonances are poles of the $S$ matrix.
Furthermore it is very suitable to implement a relativistic formalism since
the $S$-matrix elements are covariant under the Poincar\'e group.

In the Dirac or interacting picture of QFT the $S$ matrix  is given by 
\begin{align}
\label{200608.3}
S_{\beta\alpha}&
=\frac{\langle \beta|e^{i\int d^4 x {\cal L}_{\rm int}} |\alpha\rangle}{\langle 0|e^{i\int d^4 x {\cal L}_{\rm int}}|0\rangle} ~,
\end{align}
where ${\cal L}_{\rm int}$ is the interacting Lagrangian, 
$|\alpha\ra$ and $|\beta\ra$ are free particles states and
$ |0\rangle$ is the $0_{\rm th}$-order perturbative vacuum.
In Eq.~\eqref{200608.3}  $U(+\infty,-\infty)=\exp i\int d^4 x {\cal L}_{\rm int}(x)$ is the evolution operator in the
Dirac picture  from/to asymptotic times.
The denominator is a normalization factor that cancels the disconnected contributions
without involving any external particle in the matrix elements of $U(+\infty,-\infty)$.

A crucial point is that the $S$ matrix is unitary because of the completeness relation of either the
`in' or `out' states. However, it is important to emphasize that in the case of the $S$ matrix its unitarity
refers to the subset of states that are open for a given energy.
This is different to the typical sum over intermediate  states covering a resolution of the
identity for the whole Fock space.
E.g. within ordinary Quantum Mechanics (conserving the number of particles)
one can insert a resolution of the identity by plane waves 
within the product  $A B$ of two one-particle operators  as
\begin{align}
\langle \beta|AB|\alpha\rangle &=\int\frac{d^3p}{(2\pi)^3}\langle \beta| A|\vp\rangle\langle \vp|B|\alpha\rangle~.
\end{align}
Here $\vp$ takes any value, so that its kinetic energy is arbitrary large and not constrained by the
available energy fixed by the external states $|\alpha\rangle$ and $|\beta\rangle$.

After this qualification, we can show that $S^\dagger S$ is unitary by employing the completeness relation
associated with the `out' states, so that  
\begin{align}
\label{200608.2}
\int d\beta S_{\beta\gamma}^* S_{\beta\alpha}=
\int d\beta \, {{_{\rm out}\la}\beta|\gamma{\ra_{\rm in}}\!\!\!\!\!^*} \,\,\, {_{\rm out}\la}\beta|\alpha{\ra_{\rm in}}
={_{\rm in}\la}\gamma|\alpha{\ra_{\rm in}}=\delta_{\gamma\alpha}~.
\end{align}
Analogously, we can also derive $S S^\dagger=I$ by attending to the completeness relation of the
`in' states. Therefore, 
\begin{align}
\label{200220.1}
SS^\dagger=S^\dagger S=I~.
\end{align}

The scattering operator $T$, also called the $T$ matrix, is introduced such that   in terms of it the $S$ matrix reads 
\begin{align}
\label{200220.2}
S&=I+iT~.
\end{align}
The unitarity of the $S$ matrix implies in turn that $T$ fulfills that 
\begin{align}
\label{200304.1}
T-T^\dagger&=iT^\dagger T=iTT^\dagger~,
\end{align}
which is the unitarity relation for the $T$ matrix.
The last expression on the right-hand side (rhs) of the previous equation allows to derive  the Boltzmann $H$-theorem
in Statistical Mechanics, which is one of the most fundamental theorem in physics.  
It drives to the increase of entropy with time until the equilibrium is reached.
It is also well-known that unitarity implies the optical theorem and the existence of the
diffraction peak at high energies. For derivations of these points the reader can consult Sec.~3.6 of
Ref.~\cite{weinberg.vol1}.

The unitarity relation satisfied by the $T$ matrix is central in the $S$-matrix theory 
in which the scattering amplitudes are analytically continued in their kinematical arguments \cite{martin.290916.1}. 
In the development of this program one also employs the property of hermitian analyticity, 
so that the matrix elements of $T^\dagger$ can be also expressed in terms of those of $T$ by
an analytical continuation in the complex $s$ plane  of the (sub)process in question.
This fact allows an extension of the standard unitarity relation of Eq.~\eqref{200304.1}, such that 
its left-hand side (lhs) provides the discontinuity of the analytical scattering amplitudes across
the normal cuts due to intermediate states. This discontinuity implies the existence of the so-called
right-hand cut (RHC), or unitarity cut, in the scattering amplitudes. 

Due to the hermitian analyticity  the unitarity relation could also  involve on-shell intermediate states,
because the total energy is above their thresholds, but with some other kinematical variables 
taking non-physical values (e.g. the Mandelstam variable $t$ could be away from the physical process).
For more details the reader can consult Sec.~4.6 of Ref.~\cite{olive.book}.
An explicit example is developed in Sec.~\ref{sec.200522.1}, where an analytical extrapolation
in the mass of the $\eta$ squared is used for the KT formalism.  

Multiplying 
both sides of Eq.~\eqref{200304.1} 
to the left by $T^{-1}$ and to the right by ${T^\dagger}^{-1}$ we have
the interesting equation
\begin{align}
\label{200304.2b}
{T^\dagger}^{-1}-T^{-1}&=i I~.
\end{align}

The unitarity constraints are more easily expressed in terms of partial-wave amplitudes (PWAs),
in which the matrix elements of the $T$-matrix are taking between asymptotic states having well-defined
angular momentum. For instance, for two particles without spin, like in $\pi\pi$ scattering, the PWAs $T_\ell(s)$ are
giving by the standard expression
\begin{align}
T_\ell(s)&=\frac{1}{2}\int_{-1}^{+1}d\!\cos\theta\, P_\ell(\cos\theta)T(\vp',\vp)~,
\end{align}
where $\ell$ is the angular momentum,
$\vp'$ and $\vp$ are the final and initial three-momenta, $\theta$ is their relative angle
and $P_\ell(\cos\theta)$ is a Legendre polynomial. 
The general formulas relating the PWAs and the scattering amplitudes can be
found in Refs.~\cite{oller.book,oller.review}, to which we refer for further details.
Ref.~\cite{martin.290916.1} offers a rather thorough treatment on PWAs  within the helicity formalism. 

Because of time-reversal symmetry the $T$ matrix is symmetric in partial waves.
If we write this matrix in brief  as $T_L$, and denote its matrix elements by $T_{L,ij}$, by its symmetric
character we mean that $T_{L,ij}(s)=T_{L,ji}(s)$.
Eq.~\eqref{200304.2b} then implies that the imaginary part of the inverse of the PWA matrix  is fixed by unitarity. 
In the region of energy in which the resolution of the identity is saturated by two-body intermediate states,
the rhs of Eq.~\eqref{200304.2b} can be written as
\begin{align}
\label{200304.3b}
\Im T_L^{-1}&=-\frac{q}{8\pi\sqrt{s}}\theta(s)~.
\end{align}
In this equation, $q$ is the diagonal matrix of the CM three-momentum for every two-body intermediate state and
$\theta(s)$ is also another diagonal matrix whose matrix elements are 1 for $\sqrt{s}$ larger than the threshold and 0 otherwise.
Eq.~\eqref{200304.3b} is equivalent to the probably more familiar unitarity equation for PWAs 
\begin{align}
\label{200304.3c}
\Im T_L&=T_L^* \frac{q}{8\pi\sqrt{s}}\theta(s) T_L~.
\end{align}
The phase-space diagonal matrix $q(s) \theta(s)/(8\pi\sqrt{s})$ is sometimes denoted for short by $\rho(s)$.

The previous relation is not linear because its rhs is quadratic.
This fact drives to the concept of perturbative unitarity, which applies when perturbation theory is employed to calculate
the PWAs up to some order in a dimensionless parameter, let us call it $\vep$.
Therefore, if the PWA is calculated up to ${\cal O}(\vep^n)$, 
Eq.~\eqref{200304.3c} indeed implies that
\begin{align}
\label{200611.1}
\Im T_L^{(n)}&\neq T_L^{(n)*} \frac{q}{8\pi\sqrt{s}}\theta(s) T_L^{(n)}~,
\end{align}
because the rhs contains contributions of ${\cal O}(\vep^{2n})$, while the lhs only does so 
up to ${\cal O}(\vep^n)$. The consistent procedure is to expand the rhs in powers of
$\vep$ and keep only terms up to ${\cal O}(\vep^n)$. For instance, up to second
order in $\vep$ one has that 
\begin{align}
\label{200611.2}
\Im T_L^{(2)}&=T_L^{(1)*} \frac{q}{8\pi\sqrt{s}}\theta(s) T_L^{(1)}~.
\end{align}

The discontinuity across intermediate states in the crossed channels gives rise to the  crossed-channel cuts in the PWAs after
the angular projection required to calculate them. We denote this kind of cuts generically as left-hand cuts (LHCs). 
The interested reader could consult the Sec.~2 of Ref.~\cite{oller.review} for a handy pedagogical introduction
to the notions of RHC, LHC and crossing. 

Now, if we consider simultaneously stronger and weaker interactions,\footnote{Where the weaker interactions,
which are supposed to be proportional to some small dimensionless parameter, could
correspond e.g. to actually electromagnetic or weak probes, while the stronger ones typically refer to the
strong interactions among hadrons.}   the unitarity relation, at leading order in the weaker interaction, reads 
\begin{align}
\label{200220.3}
F-F^\dagger=iT^\dagger F~.
\end{align}
In this equation $F$ represents the matrix elements of the $T$ matrix involving the weaker
interactions, so that they vanish if these interactions are neglected altogether, while still the stronger ones 
would be acting.
In the latter equation we have taken that the weaker interactions act in the initial state, otherwise
write $iF^\dagger T$ on the rhs of Eq.~\eqref{200220.3}.

In PWAs the unitarity relation of Eq.~\eqref{200220.3} gets its simplest form.
In the physical region for the reactions to occur it reads
\begin{align}
\label{200220.4}
\Im F_i(s)&=\sum_jF_j(s)\rho_j(s)T_{L,ij}(s)^*=\sum_jF_j(s)^*\rho_j(s)T_{L,ij}(s)~,
\end{align}
where $\rho_j(s)$ corresponds to the phase space of the
intermediate hadronic states (integrations could also be involved for multiparticle states) and $s$ is the standard Mandelstam variable
corresponding to the total CM energy squared. The opening of the threshold for the channel $j$,
$s_{\zh,j}$,  is accounted for by a Heaviside function $ \theta(s-s_{\zh,j})$ included as part of $\rho_j(s)$.

For the one-channel case the sum on the rhs of the Eq.~\eqref{200220.4} collapses to just one term,
\begin{align}
\label{200220.5}
\Im F(s)&=F(s)\rho(s)T_\ell(s)^*=F(s)^*\rho T_\ell(s)~,
\end{align}
where  $T_\ell(s)$ is the corresponding uncoupled PWA.
Since the lhs of the equation is real then
it follows  that the phase of the form factor $F(s)$ and the phase shift of $T_\ell(s)$ are the same modulo $\pi$.
This is the well-known Watson final-state theorem.

\section{ERE, $K$-matrix, IAM and Pad\'e approximants}
\label{sec.200511.1}
\setcounter{equation}{0}   

Along this section we follow a multifaceted discussion relating different unitarization approaches, like the
(generalized) ERE, $K$-matrix parameterizations, the IAM and
the Pad\'e approximants. 

\subsection{ERE and $K$-matrix approaches}
\label{sec.200523.1}

In the early days of PCAC, soft pions theorems and realizations  based on chiral Lagrangians, it was
customary to refer as (generalized) ERE to a unitarization method based on the
identification of a remnant in the inverse of a PWA free of RHC which was expanded in powers of $p^2$.
The standard ERE was originally derived in Ref.~\cite{bethe.ere} for $NN$ interactions which,  for an uncoupled PWA, has the form
\begin{align}
\label{200512.1}
T_\ell&=\frac{p^{2\ell}}{p^{2\ell+1} \cot\delta_\ell-ip^{2\ell+1} }~.
\end{align}
The remaining part is identified with $p^{2\ell+1}\cot\delta_\ell$ as it is well known, because of the
relation between the $T$ and $S$ matrices in the normalization used typically for the ERE, which is the
one in Eq.~\eqref{200512.1}. Namely, the steps are
\begin{align}
\label{200512.2}
S&=e^{2i\delta_\ell}=1+i2 p T_\ell \rightarrow T_\ell=\frac{e^{2i\delta_\ell}-1}{2ip}~,\\
T_\ell^{-1}&=
ip\frac{e^{2i\delta_\ell}+1}{e^{2i\delta_\ell}-1}-ip=p\cot\delta_\ell-ip~.\nn
\end{align}
The $NN$ scattering is non-relativistic (NR), with $m^2\gg p^2$ at low energies, so that the expansion of
$p^{2\ell+1}\cot \delta_{\ell}$ is a Taylor series in $p^2$. 
However, for pion-pion interactions, where $p^2\sim m_\pi^2$ in the
region of interest both theoretical and experimentally speaking, the series expansion in $p^2$ is a
Laurent series for the $S$ waves. The reason is  because the Adler zeroes required by chiral symmetry
in the $S$-wave PWAs \cite{adler.181115.1}, despite there is no centrifugal barrier.
The latter is present for the higher partial waves, $\ell\geq 1$,
which implies the standard zero at threshold so that $T_\ell$ vanishes as $p^{2\ell}$ for $p\to 0$.

The phase space factor for relativistic systems changes in comparison with the NR expression of
Eq.~\eqref{200512.1}. The steps are the same as in Eqs.~\eqref{200512.1} and \eqref{200512.2}, but now
instead of $T_\ell$ one should use $T_\ell/\sqrt{s}$ so that $S_\ell=1+2ipT_\ell/\sqrt{s}$. Then,
\begin{align}
\label{200512.2b}
T_\ell&=\left[\frac{p}{\sqrt{s}}\cot\delta_\ell-i\frac{p}{\sqrt{s}}\right]^{-1}~.
\end{align}
In more recent times, the remaining part of $T_\ell^{-1}$ after discounting the factor $-ip/\sqrt{s}$, required by unitarity,
cf. Eq.~\eqref{200304.3b},
is called the inverse of the $K$-matrix, $K_\ell$, instead of $p\cot\delta_\ell$.
In this notation, $T_\ell$ is written as
\begin{align}
\label{200513.1}
T_\ell&=\left[K_\ell^{-1}-i\frac{p}{\sqrt{s}}\right]^{-1}~.
\end{align}
Of course, Eqs.~\eqref{200512.2}, \eqref{200512.2b} and \eqref{200513.1} can be generalized straightforwardly to a matrix notation for coupled-channel scattering, 
with $T_\ell$ and $K_\ell$ replaced by the matrices $T_L$ and $K_L$, respectively. The inverse of the later is usually referred
as the $M_L$ matrix, $M_L=K_L^{-1}$ \cite{au}. 

We are surprised that in these first works, e.g.
\cite{schnitzer.200221.1,schnitzer.200302.1,brown.200302.1,sakurai,chew,lehmann.200302.1,brehm}, 
it was common to refer to the (generalized)  ERE without any mention at all to the $K$-matrix approach, a notion much more common
in later times and, in particular, for more recent papers based on the  unitarization of ChPT.  
Probably this is related to the fact that the $K$-matrix parameterizations
have been used in many instances in the literature over large energy intervals in order to fit experimental data. 
As a result, it does not really make sense to keep
any memory of a particular threshold, as it is the case for the ERE.
Indeed, in those earlier papers referred the basic object of study was $\pi\pi$ scattering or the $\pi$ vector form factor. 

Another fact worth stressing is that in those earlier references the expressions 
finally used for $T_L^{-1}$ had better analytical properties than the ones typically
found later in papers using the $K$-matrix approach, as in Refs.~\cite{au,sarasa,moir} among many others
phenomenological studies.
The reason is because the later ones keep only the term $-ip/\sqrt{s}$ in $T_L^{-1}$ while, in the 
first papers referred \cite{schnitzer.200221.1,schnitzer.200302.1,brown.200302.1,sakurai,chew,lehmann.200302.1,brehm},
the non-trivial analytical function $h(s)$, which is  $8\pi g(s)$ modulo a constant, cf. Eq.~\eqref{200508.2}, 
was used by performing a dispersion relation (DR) along the RHC.
Namely,
\begin{align}
\label{200514.1}
h(s)&=8\pi g(s)-\frac{1}{\pi}\log\frac{m_\pi}{\mu}=\sigma(s)\log\frac{\sigma(s)+1}{\sigma(s)-1}~.
\end{align}
The function $g(s)$ is an analytical function of $s$ in the cut complex $s$ plane, having the RHC along the real $s$
axis for $s>4m_\pi^2$.
As a trivial byproduct, the zero at $s=0$ that occurs in the phase space factor $-ip/\sqrt{s}$
in the simplest $K$-matrix parameterizations is absent when using the function $g(s)$, which is the correct analytical
extrapolation of the two-body unitarity requirement above threshold. 
Indeed, the removal of this spurious singularity at $s=0$ was the argument used in Ref.~\cite{brown.200302.1} 
to construct the function $h(s)$ without using any DR. This reference also notices the presence of the Adler zeros
in the $I=0,$~2 S-wave $\pi\pi$ PWAs and similar expressions to Eq.~\eqref{200509.3} are proposed for these PWAs.
The main difference, an important one indeed, between Eq.~\eqref{200509.3} and Ref.~\cite{brown.200302.1}
is that the function $g(s)$,   contrary to $h(s)$, contains a subtraction constant 
\begin{align}
\label{200513.2}
\frac{1}{16\pi^2}\log\frac{m_\pi^2}{\mu^2}~,
\end{align}
which is absent in the function $h(s)$ of Brown and Gobble \cite{brown.200302.1}.
This is a crucial fact for the right reproduction of important features in low-energy $\pi\pi$ scattering,
like the generation of the $f_0(500)$ resonance pole in good agreement with the latest
and more sophisticated determinations \cite{pdg}.
As a matter of fact, the predicted $I=J=0$ $\pi\pi$ phase shifts in Ref.~\cite{brown.200302.1} are around a factor 2
smaller than data for the energies in between $500-700$~MeV, while the $I=2$ S-wave $\pi\pi$
phase shifts are too large in modulus by the same factor. These deficiencies in the approach of Ref.~\cite{brown.200302.1}
are cured once the subtraction constant
of Eq.~\eqref{200513.2}, with a natural value for $\mu\simeq 1$~GeV, is taken into account \cite{npa}.

For the $I=J=1$ $\pi\pi$ PWA Ref.~\cite{brown.200302.1} performs a generalized ERE up to and including the effective range,
\begin{align}
\label{200513.3}
T_{11}^{-1}-h(s)&=\frac{1}{a_1p^2}+\frac{r_1}{2}~.
\end{align}
The parameter $a_1$ is fixed from the current algebra prediction \cite{weinberg.200302.1},
$a_1=1/12\pi f_\pi^2$, while $r_1$ is determined by the vanishing of the real part of $T_{11}(s)^{-1}$ at $s=m_\rho^2$.
The resulting equation is therefore, 
\begin{align}
\frac{1/a_1}{m_\rho^2/4-m_\pi^2}+\frac{r_1}{2}+\Re h(m_\rho^2)=0~.
\end{align}
Let us notice that $r_1/2$ in Eq.~\eqref{200513.3} can be also considered as a subtraction constant
of $g(s)$. Attending to Eq.~\eqref{200514.1} the relation is
\begin{align}
\label{140514.2}
\log\frac{m_\pi^2}{\mu^2}=\pi r\simeq -\frac{96\pi^2f_\pi^2}{m_\rho^2-4m_\pi^2}+\delta r~,
\end{align}
with $\delta r$ a correction of around a 20\% of the term explicitly shown.
This simple calculation illustrates the discussion at the Introduction regarding the
huge unnatural value $\mu\simeq 1.7~$TeV that results by the matching in Eq.~\eqref{140514.2},
while the expected value is around 1~GeV.

As a result of this analysis the authors of Ref.\cite{brown.200302.1}
predicted the width of the $\rho(770)$ to be 130~MeV and the
$I=J=1$ phase shifts up to 1000~MeV, in good shape compared with later experimental determinations.
They also gave an expression for the coupling of the $\rho\to\pi\pi$ ($g_{\rho\pi\pi}$)  
in terms of $f_\pi$ and $m_\rho$,
which drives to the KSFR relation \cite{ksfr}, $f_\rho^2=m_\rho^2/2f_\pi^2$,
if one assumes vector-meson dominance (VMD) \cite{vmd1,vmd2}.
Here $f_\rho$ is the coupling of the $\rho$-photon transition which is equal to $g_{\rho\pi\pi}$ within VMD \cite{vmd2}.

The authors summarize their research by stating that
the fulfillment of the low-energy current-algebra constraints together with
the inclusion of extra energy
dependence as required by general principles, such as it follows by implementing 
two-body unitarity and the correct analytical properties of PWAs, are able to provide good results in
a large energy range, much larger than the one naively expected for current-algebra results.
This is a conclusion that has been strengthened along the years, at the same time that the chiral calculations
have been improved going to higher orders and the unitarization methods have become more
sophisticated.

\subsection{ERE and IAM}
\label{sec.200523.2}
 Already at 1972 the calculation of the NLO ChPT amplitude was worked by Lehmann \cite{lehmann.200302.1} in the
chiral limit ($m_\pi\to 0$), much earlier than the seminal paper by Gasser and Leutwyler
\cite{leutwyler.200220.1},  which established the general framework for ChPT at ${\cal O}(p^4)$.
The author did not need to work out the chiral Lagrangians at NLO order because he only used unitarity, crossing
symmetry and analyticity to work out the chiral loops. The point is that because of unitarity a PWA satisfies
Eq.~\eqref{200304.3c}. However, unitarity is only satisfied perturbatively in the chiral expansion,
so that if we denote by $T_4(s)$ a one-loop ChPT PWA and $T_2(s)$ its LO, then perturbative unitarity requires that
\begin{align}
\label{200514.6a}
\Im T_4(s)&=T_2(s)^2 \frac{p}{8\pi\sqrt{s}}\theta(s-4m_\pi^2) ~, 
\end{align}
a particular example of Eq.~\eqref{200611.2}.

The PWA $T_4(s)$ has LHC and RHC.
The discontinuity along the RHC is twice $i\Im T_4(s)$, because of the
Schwarz reflection principle. A DR that results by considering a closed circuit engulfing the RHC,
implies the following contribution to $T_4(s)$,
\begin{align}
\label{200514.6}
a+bs+cs^2
+\frac{s^3}{8\pi^2}\int_{4m_\pi^2}^\infty ds'\frac{T_2(s')^2\sqrt{s'/4-m_\pi^2}/\sqrt{s'}}{(s')^3(s'-s)}~.
\end{align}
Three subtractions have been taken because $T_2(s)$ at most diverges like $s$ in the limit $s\to \infty$.
By invoking crossing one can build up the one-loop contributions from the $t$- and $u$-channels for a given
process. As usual the Mandelstam variables are indicated by $s$, $t$ and $u$ ($s+t+u=0$ for massless pions).

In Cartesian coordinates for the pions and treating all of them on equal footing,
so that they are all e.g. incoming, one can write for the scattering amplitude 
$\pi_1(k_1)\pi(k_2)\pi_{i_3}(k_3)\pi_{i_4}(k_4)\to 0$,
where the $k_i$ are the on-shell four-momenta ($k_i^2=0$, $\sum_i k_i=0$), the expression
\begin{align}
\label{200515.1}
T(s,t,u)&=\delta_{i_1i_2}\delta_{i_3i_4}A(s,t,u)+\delta_{i_1i_3}\delta_{i_2i_4}A(t,s,u)+\delta_{i_1i_4}\delta_{i_2i_3}A(u,t,s)~.
\end{align}
Here crossing has also been used to properly exchange the arguments of the $A(s,t,u)$ function.
The previous expression is manifestly symmetric in the indices $i_3$ and $i_4$ which also implies that, because
the pions are bosons, $A(s,t,u)$ is symmetric under the exchange $t\leftrightarrow u$. 
Since the isospin coordinates run only from 1 to 3, two out of the four pions have the same coordinates necessarily.

In the calculation of Ref.~\cite{lehmann.200302.1} the resulting expression  for $A(s,t,u)$ has two parts. 
One of them corresponds to DR integrals of the type in Eq.~\eqref{200514.6},
in all the $s$-, $t$- and $u$-channels, which can be evaluated in an algebraic close form.
The other contribution is a second-order polynomial in the Mandelstam variables,  
whose general expression can be written as $a+bs+cs^2+c'(t^2+u^2)$, which can also be extra constrained.  
In this respect, $a=0$ because Goldstone particles do not interact
in the limit in which masses and four-momenta vanish.
The term $b s$ is order $p^2$ and it is already accounted for in $T_2(s)$.
As a result, the one-loop calculation of Lehmann only involves two unknown parameters,
nowadays typically called counterterms because they are associated to bare parameters appearing
at the NLO ChPT Lagrangian. 

In terms of the $A(s,t,u)$ amplitude one can calculate the different $\pi\pi$ isospin PWAs \cite{iam.oller.long}, $T_{IJ}$.
An interesting point of Ref.~\cite{lehmann.200302.1} is the perturbative matching in the chiral expansion
of the calculated PWAs at ${\cal O}(p^4)$ with the ERE expression for a PWA, cf. Eq.~\eqref{200512.2}.
The subtle point is that the former only satisfies unitarity in a perturbative way, as discussed above.
Therefore, writing in the massless case  that
\begin{align}
\label{200515.2}
\frac{\Re T_{IJ}}{\Im T_{IJ}}&=\cot\delta_{IJ}~,
\end{align}
is not right. The correct procedure is to write a chiral expansion of $1/T_{IJ}$ up to NLO and from there
to identify $\cot\delta_{IJ}$,
\begin{align}
\label{200515.3}
\frac{1}{T_{IJ}}&=\frac{1}{T_2+T_4}+{\cal O}(p^6)=\frac{1}{T_2}-\frac{T_4}{T_2^2}+{\cal O}(p^6)~.
\end{align}
Taking into account the perturbative unitarity satisfied by $T_4$, one can extract from here the NLO expression
for $\cot\delta_{IJ}$ (with a numerical normalization factor properly chosen) as, cf. Eq.~\eqref{200512.2b}, 
\begin{align}
\label{200515.4}
\frac{p}{\sqrt{s}}\cot\delta_{IJ}&=\frac{1}{T_2}-\frac{\Re T_4}{T_2^2}+{\cal O}(p^6)~.
\end{align}
This is indeed the first example that we know of a paper in the literature deriving the
expression of a PWA as
\begin{align}
\label{200515.5}
T_{IJ}&=\frac{T_2^2}{T_2-T_4}~.
\end{align}
This formula, generalized to any other two-body PWA and also to coupled channels,
is the basic one for the so-called IAM \cite{dobado.90,iam.oller.long}.
It also illustrates the connection between these earlier treatments based on the ERE and
this more modern method, which was named IAM after the general framework for the one-loop calculations
in ChPT was established in Ref.~\cite{leutwyler.200220.1}. The approach of Ref.~\cite{lehmann.200302.1} has
the advantage over the previous ERE of
Refs.~\cite{schnitzer.200221.1,schnitzer.200302.1,brown.200302.1,sakurai,brehm} that chiral one-loop contributions in the crossed channels
are also kept, so that the LHC is reproduced up to NLO in the inverse of the PWA. 

The extension of Eq.~\eqref{200515.5} up to two-loop ChPT can be done straightforwardly by expanding
the inverse of $(T_2+T_4+T_6)^{-1}$ up to next-to-next-to-leading order (NNLO), or ${\cal O}(p^2)$.
The result is,
\begin{align}
\label{200515.5b}
T_{IJ}&=\frac{T_2^3}{T_2^2-T_4 T_2+T_4^2-T_2 T_6}~.
\end{align}
Taking into account that perturbative unitarity requires that $\Im t_6=2 T_2\rho \Re T_4$, it follows that
$T_{IJ}$ given by Eq.~\eqref{200515.5b} fulfills exact unitarity, $\Im T_{IJ}^{-1}=-\rho$.
The Eq.~\eqref{200515.5b} is the IAM at the two-loop order \cite{arriola.2iam}.

\subsection{IAM and Pad\'e approximants}
\label{sec.200523.3}

Another non-perturbative method used with the aim of improving the convergence of the QFT
calculations in perturbation theory is the
Pad\'e resummation technique \cite{basdevant.zinn}.
It is also a unitarization method that
was applied since the early days of current algebra calculations by Refs.~\cite{basdevant.lee,basdevant.review}, in which the linear
$\sigma$ model was considered too. An interesting {\it qualitative} agreement with data
for the $\pi\pi$ $S$-, $P$- and $D$-waves was reported, despite the limitations of the theoretical input. 

Given a function $f(z)$ that is analytic at $z=0$, its Taylor series expansion around this point
converges within the circle of radius $R$, which  is the distance to the nearest singularity.
However, it is also known that the value of $f(z)$ at a point $z_1$ within its domain of analyticity,
but beyond the radius of convergence of the Taylor series around $z=0$, is fixed
by the coefficients in the later expansion.
The idea of the Pad\'e method is to provide a resummation of the Taylor series
and build an approximation of $f(z)$ beyond the radius of convergence of its Taylor series around $z=0$.

The Pad\'e approximant $[n,m]$ is given by the ratio of two polynomial functions $P_n(z)$ and $Q_m(z)$
of degrees $n$ and $m$, respectively, which has the same $n+m$ first derivatives as $f(z)$ at $z=0$. Namely,
\begin{align}
\label{200516.1}
f^{[n,m]}(z)&=\frac{P_n(z)}{Q_m(z)}=f(z)+{\cal O}(z^{n+m+1})~,~|z|<R~.
\end{align}
Notice that in particular the approximant $[n,0]$ is identical up
to ${\cal O}(z^n)$ with the Taylor series of $f(z)$ at $z=0$. 
It is also typically the case that the Pad\'e approximants usually provide an acceleration in the rate of convergence
of the Taylor series itself. For instance, one can write  that 
\begin{align}
\label{200516.2}
\sqrt{1+z}&=1+\frac{z}{1+\sqrt{1+z}}~.
\end{align}
By iteration it can be expressed as a continued fraction, which are particular cases of Pad\'e approximants,
\begin{align}
\label{200516.3}
f^{[1,0]}&=\frac{2+z}{2}~,\\ 
f^{[1,1]}&=\frac{4+3z}{4+z}~,\nn\\ 
f^{[2,1]}&=\frac{8+8z+z^2}{8+4z}~,\nn\\
f^{[2,2]}&=\frac{16+20z+5z^2}{16+12z+z^2}~,\nn
\end{align}
etc. Let us compare the first four Pad\'e approximants with the first four terms in the Taylor series,
$\sqrt{1+z}=1+\frac{z}{2}-\frac{z^2}{8}+\frac{z^3}{16}+\ldots$ by
calculating $\sqrt{2}=1.4142$. 
We then obtain the sequence of approximate results from the Pad\'e approach $\{1.5,1.4,1.4167,1.4138\}$, and the Taylor series $\{1,1.5,1.375,1.4375\}$. It is clear
the improvement in the convergence properties achieved by the Pad\'e method in this case. 

The formulas for the IAM at one- and two-loop ChPT, Eqs.~\eqref{200515.5} and \eqref{200515.5b},
respectively, can also be obtained as Pad\'e approximants, where a generic small parameter $\vep$ accounts 
for the chiral order. Formally, we then write $T_2\to \vep^2 t_2$, $T_4\to \vep^4 t_4$ and $T_6\to \vep^6 t_6$. 
The one-loop IAM is a $[1,1]$ Pad\'e approximant:
\begin{align}
\label{200517.1}
t^{[1,1]}(s)&=\frac{a_0+\vep^2 a_2}{1+\vep^2 b_2}=\vep^2 t_2+\vep^4 t_4+{\cal O}(\vep^6)~.
\end{align}
To solve this type of equation, typically found in Pad\'e approximants,
it is convenient to rewrite Eq.~\eqref{200517.1} as 
\begin{align}
\label{200517.2a}
a_0+\vep^2 a_2&=(1+\vep^2 b_2)(\vep^2 t_2+\vep^4 t_4)+{\cal O}(\vep^6)~.
\end{align}
By matching the different powers of $\vep^2$ one has that
\begin{align}
\label{200517.2}
a_0&=0~,\\
a_2&=t_2~,\nn\\
b_2&=-t_4/t_2~.\nn 
\end{align}
From which it follows that
\begin{align}
\label{200517.3}
t^{[1,1]}&=\frac{T_2}{1-T_4/T_2}=\frac{T_2^2}{T_2-T_4}~.
\end{align}
For the approximant $[1,2]$ 
\begin{align}
\label{200517.4}
t^{[1,2]}(s)&=\frac{a_0+\vep^2 a_2}{1+\vep^2 b_2+\vep^4 b_4}
=\vep^2 t_2+\vep^4 t_4+\vep^6 t_6+{\cal O}(\vep^8)~.
\end{align}
The result of the matching is the same as in Eq.~\eqref{200517.2} for
$a_0$, $a_2$ and $b_2$, and the extra new parameter $b_4$ is 
\begin{align}
\label{200517.5}
b_4&=(t_4^2-t_2 t_6)/t_2^2 ~. 
\end{align}
Therefore,
\begin{align}
\label{200517.6}
t^{[1,2]}&=\frac{T_2}{1-\frac{T_4}{T_2}+\frac{T_4^2-T_2 T_6}{T_2^2}}
=\frac{T_2^3}{T_2^2-T_2 T_4+T_4^2-T_2 T_6}~,
\end{align}
as Eq.~\eqref{200515.5b}.

\section{Final-State Interactions}
\setcounter{equation}{0}   
\label{sec.200523.4}

As a canonical example of taking into account the FSI that correct the production processes
due to weaker probes because of the rescattering by the stronger interactions, we start with the 
unitarization of the vector pion form factor, $F_V(s)$, within the ERE approach of Ref.~\cite{sakurai}.
We next move to the Omn\`es solution for a form factor and also consider the scalar pion form factor, $F_S(s)$,
paying attention to a caveat in the use of an Omn\`es function that one should properly consider.  
Along the discussion we introduce the way FSI are treated in Ref.~\cite{truong.200211.2}, as it is
probably the first paper in which NLO ChPT is unitarized to account for FSI following the basic notions of
unitarity, Watson final-state theorem and use of an Omn\`es function, which are the basic elements usually employed in
the different modern approaches to resum FSI \cite{oller.book,oller.review}.
We end this section with a basic account of the Khuri-Treiman approach for $\eta\to 3\pi$ decays.

\subsection{ERE, the Omn\`es solution and coupled channels}
\label{sec.200514.1}

The application of the ERE  for implementing the FSI of the pion vector form factor was pioneered in Ref.~\cite{sakurai}.
The main aim of this paper concerns the  corrections because of the finite width of the $\rho$
to the VMD dominance relation between $\Gamma(\rho\to e^+e^-)$ and $\Gamma(\rho\to\pi^+\pi^-)$,
as well as to characterize the energy shape of $\Gamma(e^+e^-\to \pi^+\pi^-)$.

Ref.~\cite{sakurai} implemented the relationship between the $I=J=1$ $\pi\pi$ PWA and
the pion  form factor $F_V(s)$ by writing $F_V(s)=T_1(s)/t_2(s)$, with $t_2(s)$  the LO ChPT amplitude. 
This relation is a consequence of the Omn\`es representation in the approximation in which:
i) One assumes that the only zero in $T_1(s)$ in the region of interest
is the one at threshold, $s=4m_\pi^2$, because of the $\ell=1$ centrifugal barrier; 
ii) one also assumes the dominance of the $\rho(770)$ exchange so that it is a good approximation to 
consider that $T_1(s)$ is dominated by $s$-channel dynamics.
\footnote{Under these  assumptions $T_1(s)$ is given by the Omn\`es function on the rhs of Eq.~\eqref{200216.4}
times $(s-4m_\pi^2)/48\pi f_\pi^2$.} 
Thus,
\begin{align}
\label{200514.3}
F_V(s)&=\frac{48\pi f_\pi^2 T_1(s)}{s-4m_\pi^2}=\frac{T_1(s) a_1}{p^2}~, 
\end{align}
guaranteeing that $F_V(0)=1$ because of conservation of total charge. 
Next, Ref.~\cite{sakurai} performs the same ERE of Ref.~\cite{brown.200302.1},
which we have already discussed, cf. Eq.~\eqref{200513.3},
which allows to finally write the form factor in a successful manner as
\begin{align}
\label{200514.4}
F_V(s)=\frac{1}{1+\frac{r_1p^2}{2a_1}+\frac{p^2}{a_1}h(s)}~.
\end{align}
The authors of Ref.~\cite{sakurai} simplify further this expression  by removing those terms involving the expansion of
the real part of $h(s)$ around $s=m_\rho^2$ that are at least of ${\cal O}(s-m_\rho^2)$.
They finally write
\begin{align}
\label{200514.5}
F_V(s)&=\frac{m_\rho^2[1+d_1 m_\rho/\Gamma_\rho]}{m_\rho^2-s-im_\rho \Gamma_\rho (p/p_\rho)^3(m_\rho/\sqrt{s})}~,\\
d_1&=\frac{3}{\pi}\frac{m_\pi^2}{p_\rho^2}\log\left(\frac{m_\rho+2p_\rho}{2m_\pi}\right)
+\frac{m_\rho}{2\pi p_\rho}-\frac{m_\pi^2m_\rho}{\pi p_\rho^3}~,\nn\\
p_\rho&=\sqrt{m_\rho^2/4-m_\pi^2}~.\nn
\end{align}
Again, one concludes that the extrapolation of the current-algebra results plus the extra energy dependence that
arises by implementing  the basic principles
of two-body unitarity and analyticity allows one to reach much higher energies than
expected, even above the 1~GeV frontier.

Writing a form factor proportional to a given PWA is usually employed
in many cases in the literature. The basic reason is to provide an expression for the
coupled form factors $F_i(s)$ that automatically satisfies the constraint imposed by the two-body
unitarity, cf. Eq.~\eqref{200220.4}. Following Ref.~\cite{au} one then writes
\begin{align}
\label{200515.6}
F_i&=\sum_j \widetilde{\alpha}_j T_{ji}~,
\end{align}
where the sum is over the strongly-coupled channels.
The functions $\widetilde{\alpha}_i$ are real and they are also expected to be smooth because
all the RHC features in $F_i(s)$ are included in the PWAs $T_{ij}(s)$. As a result, the
$\widetilde{\alpha}_i$ should not have nearby singularities, if any. They could involve crossed-channel
cuts which could be mimicked typically by parameterizing these functions by low-degree polynomials.
Nonetheless, in the case of the low-energy interactions of the lightest pseudoscalars, like pions,
an extra feature is the presence of the Adler zeroes in the $S$ waves. In particular, for
$I=J=0$ we have already discussed that this Adler zero is around $s_A=m_\pi^2/2$, cf. Eq.~\eqref{200509.3}.
The existence of Adler zeros is a characteristic feature of the interactions of the Goldstone bosons, as said, but not
necessarily for their production through external currents. To handle with such cases Ref.~\cite{au}
proposes to explicitly remove the Adler zeroes in the $T_{ij}(s)$, when they are present, and any 
necessary zero in the production process is then explicitly included in the prefactors. Denoting by
$\cT_{ij}(s)=T_{ij}(s)/(s-s_{A_{ij}})$, with $s_{A_{ij}}$ the Adler zero in $T_{ij}(s)$, the final expression proposed is
\begin{align}
\label{200515.7}
F_i&=\sum_j \alpha_j \cT_{ji}~.
\end{align}

 For the case of only one coupled channel, the form factor can be expressed in terms of an Omn\`es function
$\Omega(s)$. 
 Due to the Watson final-state theorem the continuous phase of the form factor $\varphi(s)$  
is the same as the phase shift $\delta(s)$ for the PWA  $T(s)$.
The Omn\`es function results by performing a DR for the logarithm of the function 
$f(s)=F(s)Q(s)/P(s)$, where $P(s)$ and $Q(s)$ are the polynomials whose only roots are the possible
zeros and poles of $F(s)$, respectively, which are assumed to be finite in number.  
The discontinuity of  $\log f(s)$ along the RHC is the discontinuity of its imaginary 
part, and it is given by $2i\varphi(s)$.
We can then write the following expression for the DR of $\omega(s)\equiv \log f(s)$,\footnote{For a more extensive discussion
on the Muskhelishvili-Omn\`es problem the reader can consult Refs.~\cite{oller.book,oller.review}.}
\begin{align}
\label{181120.8}
\omega(s)&=\sum_{i=0}^{n-1}a_i s^i+\frac{s^n}{\pi}\int_{s_{\rm th}}^\infty \frac{\varphi(s')ds'}{(s')^n(s'-s)}~,
\end{align}
where we have taken $n$ subtractions assuming that $\vh(s)$ does not diverge stronger than $s^{n-1}$ when $s\to\infty$.
The Omn\`es function $\Omega(s)$ is defined in terms of $\omega(s)$ as 
\begin{align}
\label{181120.9}
\Omega(s)=\exp{\omega(s)}~.
\end{align}
One can always normalize the Omn\`es function such that  $\Omega(0)=1$, which fixes $a_0=1$.
In this manner we always take at least one subtraction. 
It is also clear that the ratio
\begin{align}
\label{181120.10}
R(s)&=\frac{F(s)}{\Omega(s)}~,
\end{align}
is a  meromorphic function of $s$ in the first RS of the cut complex $s$ plane,
being analytic in this whole
plane if $F(s)$ has no bound states. 
As it is well known,  any analytical function in the whole complex plane is either a constant or it is unbounded,
which is then the case for $R(s)$ too under the stated assumptions. 
Therefore,   
\begin{align}
\label{181120.11}
F(s)&=R(s)\Omega(s)~,
\end{align}
diverges as much as or stronger than $\Omega(s)$ for $s\to \infty$.
The function $\omega(s)$ would have severe divergences for $s\to\infty$ if its DR required for convergence more than one subtraction.  
The reason is that if $\varphi(s)/s^{n-1}$ ($n \geq 2$) has no zero limit for $s\to \infty$, 
the DR for $\omega(s)$ would be affected by logarithmic divergences like $s^{n-1} \log s$ which could not be  
cancelled by the subtractive polynomial. 
In such circumstances it would be required that $R(s)$ is a non-trivial analytical function in order to 
cancel such divergences and guarantee that $F(s)$ can be represented as a DR.

If the conditions are met for a DR  of $\log F(s)Q(s)/P(s)$, cf. Eq.~\eqref{181120.8}, 
then  $R(s)=Q(s)/P(s)$ is a rational function. Thus, from the previous analysis, we conclude that 
the DR of $\omega(s)$ in Eq.~\eqref{181120.8} involves only one subtraction and
it is then necessary that   $|\varphi(s)/s|<s^{-\gamma}$ for some $\gamma>0$ in the limit $s\to \infty$.  
We can then write the following representation for $F(s)$, 
\begin{align}
\label{181121.1}
F(s)&=\frac{P(s)}{Q(s)}\Omega(s)~,\\
\label{181121.2}
\Omega(s)&=\exp\omega(s)~,\\
\label{181121.3}
\omega(s)&=\frac{s}{\pi}\int_{s_{\rm th}}^\infty \frac{\vh(s')ds'}{s'(s'-s)}~.
\end{align}
The presence of $P(s)$ makes clear  that one can fix de normalization of the Omn\`es function,  $\Omega(0)=1$,
without any loss of generality. 
The asymptotic  behavior of $\Omega(s)$ in the limit $s\to \infty$ can be calculated as follows.
Let us rewrite  $\omega(s)$ in Eq.~\eqref{181121.3} as
\begin{align}
\label{181122.1}
\omega(s)&=\vh(\infty)\frac{s}{\pi}\int_{s_{\rm th}}^\infty \frac{ds'}{s'(s'-s)}
+\frac{s}{\pi}\int_{s_{\rm th}}^\infty \frac{\vh(s')-\vh(\infty)}{s'(s'-s)}ds'~, 
\end{align}
with $\vh(\infty)=\lim_{s\to\infty}\vh(s+i\ep)$. 
Then,  
\begin{align}
\label{181122.2}
\omega(s+i\ve)&\xrightarrow[s\to\infty]{} -\frac{\vh(\infty)}{\pi}\log \frac{s}{s_{\rm th}} + i\vh(\infty)
-\frac{1}{\pi}\int_{s_{\rm th}}^\infty\frac{\vh(s')-\vh(\infty)}{s'}ds'~,
\end{align}
being the limit $s\to\infty$ dominated by the logarithmic divergence,
as the other two terms in this equation are  constants.
It follows from here the limit behavior  
\begin{align}
\label{181122.3}
\Omega(s)&\xrightarrow[s\to \infty]{} {\cal C}_{\Omega}\, e^{i\vh(\infty)}\times
\left(\frac{s_{\rm th}}{s}\right)^\frac{\vh(\infty)}{\pi}~.
\end{align}
This result, together with Eq.~\eqref{181121.1}, implies that the asymptotic behavior for  $F(s)$ is 
\begin{align}
\label{181122.4}
F(s)&\xrightarrow[s\to\infty]{}{\cal C}_{F}\, e^{i\vh(\infty)}\times s^{p-q-\frac{\vh(\infty)}{\pi}}~,  
\end{align}
where $C_{\Omega}$ and $C_{F}$ are constants, and $p$ and $q$ are the number of zeros and poles of $F(s)$, respectively (or equivalently,  
the degrees of $P(s)$ and $Q(s)$, in this order).  
Two interesting consequences follow from  Eq.~\eqref{181122.4}:

\noindent
i) If the asymptotic high-energy behavior of $F(s)$ is known to be proportional to $s^\nu$, then 
\begin{align}
\label{181122.5}
p-q-\frac{\vh(\infty)}{\pi}=\nu~.
\end{align}

\noindent
ii) Under changes of the parameters  when modeling strong interactions one should keep Eq.~\eqref{181122.5} 
unchanged.  As $\nu$ is a known constant,  then
\begin{align}
\label{181122.6}
p-q-\frac{\vh(\infty)}{\pi}={\rm fixed}~.
\end{align}
For instance,  if $\vh(\infty)/\pi$ increases by one  and there are no bound states
 then an extra zero should be introduced in the form factor to satisfy Eq.~\eqref{181122.6}. 
A similar procedure would be applied for other scenarios.

It is worth stressing that by using Eq.~\eqref{181121.1} one can guarantee that  Eq.~\eqref{181122.6} is fulfilled,
while this is not the case for $\Omega(s)$.
The use of this function without taking proper care of the rational function $P(s)/Q(s)$, included in the expression for
$F(s)$ in Eq.~\eqref{181121.1},  could drive to an unstable behavior under changes of the parameters, e.g. in a fit to data.
This problem was originally discussed in Ref.~\cite{ollerz.200220.1} in connection with the scalar form factor of the pion $F_S(s)$,
to which we refer for further details in the discussion that follows. 
This form factor is associated with the light-quark scalar source, $\bar{u}u+\bar{d}d$, and is defined as
\begin{align}
\label{181122.7}
F(s)&=\hat{m}\int d^4 x e^{i(p+p')x}\langle0|\bar{u}(x)u(x)+\bar{d}(x)d(x)|0\rangle~,
\end{align}
where $u$ and $d$ are the up and down quarks,  $\hat{m}$ is their masses, and $s=(p+p')^2$.
Because of the quantum numbers of the non-strange scalar source, the FSI 
occur in the isoscalar scalar meson-meson scattering, introduced in Sec.~\ref{sec.200511.1}.
There, we discuss the Adler zero required by chiral symmetry and the pole of the 
$\sigma$ or $f_0(500)$ resonance, being both of them related by unitarity, analyticity and chiral symmetry. 
At around the two-kaon threshold, $\sqrt{s}=991.4$~MeV, the $K\bar{K}$ channel makes a big impact.
This energy almost coincides with the sharp emergence of the  $f_0(980)$ resonance,
which gives rise to a rapid increase of the $\pi\pi$  isoscalar scalar phase shifts $\delta_{00}(s)$, since
it is a relatively narrow resonance \cite{pdg}, cf. Fig.~\ref{fig.200524.1}.
The elasticity parameter $\eta_{00}$ also experiences a sharp reduction as soon as the $K\bar{K}$ channel open, 
since the $f_0(980)$ couples much more to $K\bar{K}$ than to $\pi\pi$ \cite{guo.181123.1}.
This phenomenon causes an active conversion of the pionic flux into the kaonic one.
   
The rapid rise of the isoscalar scalar $\pi\pi$ phase shifts,  
also implies the corresponding rise of the phase of the isoscalar scalar PWA $T(s)$, $\vh(s)$, because they coincide 
below the $K\bar{K}$ threshold, i.e. for $\sqrt{s}<2m_K$. 
However, above this energy the rise of $\vh(s)$ is  interrupted  abruptly
if $\delta_{00}(s_K)<\pi$, with $s_K=4m_K^2$, while in the opposite case $\vh(s)$ keeps increasing. 
Quite interestingly, the two situations can be connected by tiny variations in the values of the parameters in the
hadronic model, while keeping compatibility with the experimental phase shifts at around the $f_0(980)$ mass.

As a result, there is a jump in the limiting value of $\Omega(s)$ because $\varphi(\infty)$ changes by $\pi$.
Thus, in order to keep constant Eq.~\eqref{181122.6} under an increase by $\pi$ in $\varphi(\infty)$ for
$\delta_{00}(s_K)>\pi$, it is necessary to increase $p$ by one unit, so that a zero is necessary in $F_S(s)$ that is not
present when $\delta_{00}(s_K)<\pi$.
For completeness, we also mention that had we required the continuity from $\delta_{00}(s_K)>\pi$
to $\delta_{00}(s_K)<\pi$ then an extra pole (in the first RS) should be added.  
This latter scenario can be disregarded in $\pi\pi$ scattering because of the absence of bound states.\footnote{With 
respect to the difference between $\varphi(s)$ and $\delta_{00}(s)$, as indicated above,
the $f_0(980)$ dominates the behavior of the isoscalar scalar meson-meson 
scattering around 1~GeV, and couples much more strongly to kaons than to pions. 
For instance, Ref.~\cite{guo.181123.1} calculates that its coupling to kaons is a  factor 3 larger than that to pions.
This makes that the mixing between the pion and kaon scalar form factors is suppressed, following each of them  
its own eigenchannel of the $I=J=0$ PWAs.}

Let $s_1$ be the value of $s$ at which the pion scalar form factor has a zero for $\delta(s_K)>\pi$. Then, we can write
an Omn\`es representation of the pion scalar form factor in terms of a modified Omn\`es function
\begin{align}
\label{181123.5}
\Omega(s)&=\left\{
\begin{array}{ll}
\exp\omega(s)& ~,~\delta(s_K)<\pi~,\\
\frac{s_1-s}{s_1}\exp\omega(s) & ~,~\delta(s_K)>\pi~, 
\end{array}
\right.
\end{align}
such that $F_S(s)=F_S(0)\Omega(s)$. From here it is clear that $s_1$ can be fixed by the
requirement that $\Im F(s_1)=0$. 
Because of the Watson final-state theorem in the elastic region 
we can write that $\Im F(s)=|F(s)|\sin  \delta_{00}(s)|/\rho(s)$ and it vanishes when 
$\delta_{00}(s_1)=\pi$, which allows to determine $s_1$ from the knowledge of $\delta_{00}(s)$. 
The context clarifies whether the same symbol $\Omega(s)$ actually refers to Eq.~\eqref{181120.9} or Eq.~\eqref{181123.5}.

A clear lesson from the discussion here is that possible troubles could occur when using an Omn\`es function
in fitting the free parameters because an unstable behavior could arise due to a jump in $\vh(\infty)$.
These regions of dramatic differences in $\exp\omega(s)$ are separated by a  discontinuity of $\vh(s)$ in the parametric space. 
As a consequence, it is important in the fitting process to satisfy the condition Eq.~\eqref{181122.6}. 
In particular, for the $I=J=0$ $\pi\pi$ PWA the more elaborated function in Eq.~\eqref{181123.5} should be used, instead of just the 
standard Omn\`es $\exp\omega(s)$ given in Eq.~\eqref{181121.2}.
This fact also affects studies of two-photon fusion into two
pions, like Ref.~ \cite{pen.prl}, as discussed in Ref.~\cite{schat.gama}.

\subsection{The IAM for FSI}
\label{sec.200515.1}

The first step of Ref.~\cite{truong.200211.2} is to write down twice subtracted DR expressions for the scalar and vector pion form factors,
$F_S(s)$ and $F_V(s)$, respectively, as
\begin{align}
\label{200216.1}
F_S(s)&=1+\frac{\la r_S^2\ra s}{6}+\frac{s^2}{\pi} \int_{4m_\pi^2}^\infty
\frac{F_S(s')e^{-i\delta_{00}}\sin\delta_{00}(s')ds'}{s^{'2}(s'-s-i\ep)}~,\\
\label{200216.2}
F_V(s)&=1+\frac{\la r_V^2\ra s}{6}+\frac{s^2}{\pi} \int_{4m_\pi^2}^\infty
\frac{F_V(s')e^{-i\delta_{11}}\sin\delta_{11}(s')ds'}{s^{'2}(s'-s-i\ep)}~.
\end{align}
Here, $\delta_{00}(s)$ and $\delta_{11}(s)$ are the $J=0$ and 1 isoscalar and isovector
$\pi\pi$ phase shifts, in this order. 
These DRs can be interpreted as singular integral equations (IEs) for the form factors $F_S(s)$ and
$F_V(s)$ \cite{jamin.ff}.

Let us remark, as in   Ref.~\cite{truong.200211.2}, 
that the solutions of the IEs of Eqs.~\eqref{200216.1} and \eqref{200216.2}
for $F_S(s)$ and $F_V(s)$, respectively, can be expressed in terms of the associated Omn\`es
functions \cite{meissgasser}.
In the approximation of identifying the phases of the form factors with
the phase shifts, strictly valid only for the elastic region, one has the approximate expressions
\begin{align}
\label{200216.3}
F_S(s)&=P_S(s)\exp\left[\frac{s}{\pi}\int_{4m_\pi^2}^\infty \frac{\delta_{00}(s')ds'}{s'(s'-s-i\ep)}\right]~,\\
\label{200216.4}
F_V(s)&=P_V(s)\exp\left[\frac{s}{\pi}\int_{4m_\pi^2}^\infty \frac{\delta_{11}(s')ds'}{s'(s'-s-i\ep)}\right]~,
\end{align}
where $P_S(s)$ and $P_V(s)$ are polynomials that take into account the zeros (if any) of the form factors in the
first or physical RS.

At the one-loop order in ChPT or, equivalently, at next-to-leading order NLO or ${\cal O}(p^4)$,
we can replace inside the dispersive integrals of Eq.~\eqref{200216.1} the $\pi\pi$ scattering PWAs
at leading order, 
\begin{align}
\label{200219.1}
f_0(s)&=\sin\delta_{00}e^{i\delta_{00}}=\delta_{00}(s)+{\cal O}(p^4)=\frac{\sigma(s)}{16\pi}\frac{s-m_\pi^2/2}{f^2}+{\cal O}(p^4)~,\\
\label{200304.2}
f_1(s)&=\sin\delta_{11} e^{i\delta_{11}}=\delta_{11}(s)+{\cal O}(p^4)=\frac{\sigma(s)}{16\pi}\frac{s-4m_\pi^2}{6f^2}+{\cal O}(p^4)~.
\end{align}
The phase space function $\sigma(s)$ is defined in Eq.~\eqref{200508.2}. 
Evaluating the dispersive integral in Eq.~\eqref{200216.1} with the approximation for $f_0(s)$
of Eq.~\eqref{200219.1}, 
Ref.~\cite{truong.200211.2} of course ends with the same expression for $F_S(s)$ as the NLO ChPT \cite{leutwyler.200220.1}
result,
\begin{align}
\label{200219.3}
F_S(s)&=1+\frac{s}{6}\la r_S^2\ra - \frac{1}{16\pi^2 f^2}\left[(s-m_\pi^2/2)[h(s)-h(0)]+\frac{m_\pi^2}{2}h'(0)s\right]+{\cal O}(p^6)~.
\end{align}
The function $h(s)$ is defined in Eq.~\eqref{200514.1}.  
By proceeding in an analogous way, 
a similar expression holds for the vector form factor at this level of accuracy, ${\cal O}(p^4)$, 
\begin{align}
\label{200219.3b}
F_V(s)&=1+\frac{s}{6}\la r_V^2\ra - \frac{1}{96\pi^2 f^2}\left[(s-4m_\pi^2)[h(s)-h(0)]+ 4 m_\pi^2 h'(0)s\right]+{\cal O}(p^6)~.
\end{align}
There is an important difference between the scalar and vector form factors.
The unitarity corrections are enhanced by around a factor 6 for the former compared to the latter, because the
leading order ChPT amplitude is around a factor 6 larger, compared Eqs.~\eqref{200219.1} and \eqref{200304.2}, as
first noticed in Ref.~\cite{nd} and already discussed above in detail. 

By invoking the Watson final-state theorem, one can calculate from the  perturbative expressions
of $F_S(s)$ and $F_V(s)$ in Eqs.~\eqref{200219.3} and \eqref{200219.3b} the $\pi\pi$ phase shifts for $J=0$ and 1,
respectively. 
Nonetheless, since the form factors are calculated perturbatively one should
proceed consistently in order to extract from this perturbative information the corresponding phase shifts.
In this way,
denoting by $F_2(s)$ the LO form factors and by $F_4(s)=F_4^r(s)+iF_4^i(s)$ their NLO contributions,
with the superscripts indicating the real ($r$) and imaginary ($i$) parts, we then have for the
Watson final-state theorem:
\begin{align}
\label{200304.3}
F(s)&=
|F(s)|e^{i\phi}=\sqrt{(F_2+F_4^r)^2+(F_4^i)^2}e^{i\phi}=
F_2\sqrt{ \left( 1 + \frac{F_4^r}{F_2} \right) ^2 + \left( \frac{F_4^i}{F_2} \right) ^2} e^{i\phi}\\
&=F_2+F_4^r+i\phi F_2+{\cal O}(p^6)~,\nn
\end{align}
from where the phase $\phi$ can be extracted. 
Let us notice that
Ref.~\cite{truong.200211.2} compared directly the phase of the perturbative form factors
in Eqs.~\eqref{200219.3} and \eqref{200219.3b} with the
phase shifts of the $\pi\pi$ PWAs in its Fig.~1 and 2.  
In this respect,  it did no take account that this is not meaningful because
the Watson final-state theorem only holds perturbatively in ChPT, as explained.

\begin{figure}[H]
\begin{center}
\includegraphics[width=0.85\textwidth]{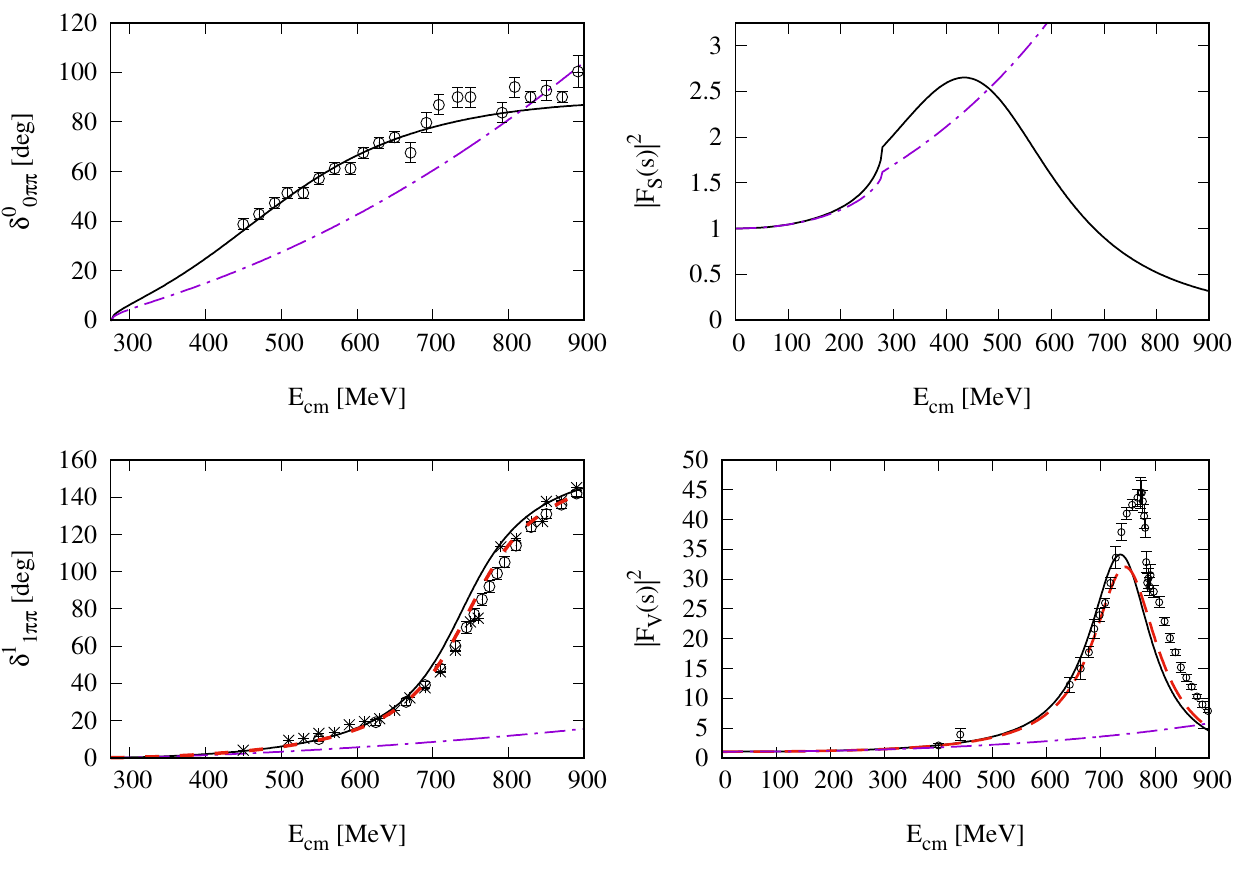} 
\caption{{\small The top row concerns  the $\pi\pi$ scalar form factor $F_S(s)$ and the
bottom one the $\pi\pi$ vector form factor $F_V(s)$.
In each row the left panel refers to the phase and the right panel to the modulus squared of the
corresponding form factor.
The perturbative calculations are indicated by the (magenta) dot-dashed lines in all cases.
The non-perturbative result for $F_S(s)$ are shown by the (black) solid lines.
For $F_V(s)$ we show two lines for the IAM resummation, Eq.~\eqref{200220.9},
the (black) solid lines and the (red) dashed ones.
The former employs $\la r_V^2\ra=0.42$~fm$^2$ (used in Ref.~\cite{truong.200211.2}) and the latter $\la r_V^2\ra=0.41$~fm$^2$.
The $\rho-\omega$ mixing, clearly visible at the top of $|F_V(s)|^2$, is not discussed here. 
The experimental points for the $I=J=1$ $\pi\pi$ phase shifts are from Refs.~\cite{pipi11}, and
those for $|F_V(s)|^2$ were obtained in Ref.~\cite{1985.barkov}.
For the $I=J=0$ phase shifts we use the subset of points  employed in Fig.~\ref{fig.200524.1}
and that appear on the top in the $f_0(500)$ region.}
\label{fig.200220.1}}
\end{center}
\end{figure}

We show in Fig.~\ref{fig.200220.1} the resulting form factors, so that the top line is dedicated to $F_S(s)$
and the bottom one to $F_V(s)$. The panels on the left correspond to the phases of these form factors and
the panels on the right to their module squared.
It is clear that there is a strong departure between the calculated phase shifts from the NLO ChPT form factors (magenta dashed lines) and the experimental values even at low values of $s$. 
This is also clearly true for the modulus squared of $F_V(s)$, for which the perturbative calculation
again departures strongly from the experimental points.
It is particularly visible there the emergence of the resonance $\rho(770)$,
which dominates the phase shifts and $|F_V(s)|^2$,
with tails extending up to threshold and affecting the low-energy results.
This phenomenon can only be captured approximately in $SU(2)$ ChPT by the large size of the counterterm $\bar{\ell}_6$, 
\begin{align}
\bar{\ell}_6&=(4\pi f_\pi)^2\la r_V^2\ra= 16.5\pm 1.1~,
\end{align}
as estimated in Ref.~\cite{leutwyler.200220.1}.

For the vector case, the cause of the large higher-order contributions  is clearly associated with the
prominent role played by the $\rho(770)$ resonance.
In turn, for the scalar sector the enhanced RHC is the one blamed for such effects.
Indeed, these strong contributions from unitarity and analyticity even drive
to the emergence of a pole in the complex $s$ plane, the $\sigma$ or $f_0(500)$ resonance, as already
discussed, cf. Eq.~\eqref{200509.3}.

 Ref.~\cite{truong.200211.2} discusses that the application of the chiral series expansion should be
performed on the inverse of the  form factor rather than on the form factor itself.
The main reason lies on sound and general grounds, as provided by unitarity and analyticity. 
Let us consider a DR representation of $F^{-1}(s)$, analogous to Eq.~\eqref{200216.1}.
The point to be stressed is that the imaginary part of $F^{-1}(s)$ is expected to be much smoother
than the imaginary part of $F(s)$ itself in the elastic region. 
The reason is that the imaginary part of the inverse of the form factor satisfies,
because of unitarity in PWAs, that 
\begin{align}
\label{200220.6}
\Im F^{-1}(s)=-\frac{\Im F(s)}{|F(s)|^2}=-\rho(s)\left(\frac{T(s)}{F(s)}\right)
=-\rho(s)\left(\frac{T(s)}{F(s)}\right)^*~.
\end{align}
As $F(s)$ and $T(s)$ share the same resonances, their propagators cancel in  
the ratio $T(s)/F(s)$ that gives $\Im F^{-1}(s)$.  
Then, this ratio is expected to be smoother than  $\Im F(s)=\rho(s)F(s)^*T(s)$, where this
cancellation does not occur but rather the resonance effects in $F(s)$ and $T(s)$ mutually enhance each other
because of the product involved.

Then, let us write down a twice-subtracted DR for the inverses of the form factors $F_S(s)$ and $F_V(s)$.
First, we neglect by now the possible presence of zeroes in the form factors in the 1st Riemann sheet (RS),
which give rise to poles in the inverse of the form factors.
The issue of a zero in $F_S(s)$ for certain types of $T$ matrices was already discussed in
Sec.~\ref{sec.200514.1}, as first shown to happen in Ref.~\cite{ollerz.200220.1}.
This is not an issue here because we are considering the one-channel elastic scattering in the
isoscalar scalar  $\pi\pi$ PWAs. 
As a result we write,
\begin{align}
\label{200220.7}
F_S^{-1}(s)&=1-s\frac{\la r_S^2\ra}{6}-\frac{s^2}{\pi}\int_{4m_\pi^2}^\infty \frac{\rho(s')e^{i\delta_{00}}\sin\delta_{00}(s)\,F_S^{-1}(s')ds'}{(s')^2(s'-s-i\ep)}~,\\
\label{200220.8}
F_V^{-1}(s)&=1-s\frac{\la r_V^2\ra}{6}-\frac{s^2}{\pi}\int_{4m_\pi^2}^\infty \frac{\rho(s')e^{i\delta_{11}}\sin\delta_{11}(s)\,F_V^{-1}(s')ds'}{(s')^2(s'-s-i\ep)}~. 
\end{align}
Then, up to ${\cal O}(p^4)$, in the integrand of these integrals one takes the leading order expressions in the
chiral expansion of  $f_\ell(s)$, cf. Eqs.~\eqref{200219.1} and \eqref{200304.2}, and $F_{S,V}(s)=1$.
In this way, except for a global sign the same result as above  is obtained for the dispersive integral as in the  DR for $F_{S,V}(s)$.
Namely, the only difference is a flip of sign in the 
NLO contributions 
in Eqs.~\eqref{200219.3} and \eqref{200219.3b}.
Then, the 
results for the form factors can be written as
\begin{align}
\label{200220.9}
F(s)&=\frac{1}{1-F_4(s)}~,
\end{align}
with $F(s)$ representing either $F_S(s)$ or $F_V(s)$ and $F_{4}(s)$ is the ${\cal O}(p^4)$ ChPT result. Similarly
$F_2(s)=1$ is the LO ChPT calculation. 
Being more specific, Eq.~\eqref{200220.9} results after performing the
DR integrals, compare with  Eqs.~\eqref{200216.1} and \eqref{200216.2}, 
\begin{align}
\label{200220.10}
F_S(s)&=\frac{1}{1-\frac{\la r_S^2\ra s}{6}-\frac{s^2}{(4\pi f_\pi)^2} \int_{4m_\pi^2}^\infty
\frac{\sigma(s')(s'-m_\pi^2)ds'}{s^{'2}(s'-s-i\ep)}}~,\\
\label{200220.11}
F_V(s)&=\frac{1}{1-\frac{\la r_V^2\ra s}{6}-\frac{s^2}{6(4\pi f_\pi)^2} \int_{4m_\pi^2}^\infty
\frac{\sigma(s')(s'-4m_\pi^2)ds'}{s^{'2}(s'-s-i\ep)}}~.
\end{align}

The resulting phase and modulus squared of $F_S(s)$
from Eq.~\eqref{200220.10} is shown by the (black) solid lines in the top panels of
Fig.~\ref{fig.200220.1}.
The resummed expression of $F_V(s)$ in Eq.~\eqref{200220.11}
gives rise to the results shown by the (black) solid and the (red) dashed lines
in the bottom panels of Fig.~\ref{fig.200220.1}. 
They differ in the value of $\la r_V^2\ra$ employed, so that the former uses
0.42~fm$^2$ (as Ref.~\cite{truong.200211.2}), 
and the latter takes the slightly lower value 0.41~fm$^2$, so as to agree better with the data on the 
isovector vector $\pi\pi$ phase shifts.
We also use the updated value $f=92.4$~MeV, instead of 94~MeV used  in Ref.~\cite{truong.200211.2}.\footnote{This reference indeed employs the normalization $f=133$~MeV=$94 \sqrt{2}$.}
It is clear that now, the resulting phase shifts calculated from the phases of the form factors in
Eqs.~\eqref{200220.10} and \eqref{200220.11}  
are much closer to the experimental points than the perturbative ones form  Eqs.~\eqref{200219.3} and \eqref{200219.3b}. 
The same dramatic improvement also happens for the modulus squared of $F_V(s)$
calculated from Eq.~\eqref{200220.11}, as compared with the data points given by the empty circles.
In the peak of $|F_V(s)|^2$ it is clear the effect due to the $\rho-\omega$ mixing, which is not
treated here, see e.g. Ref.~\cite{palomar} for its implementation. 
Notice that this improvement is achieved by employing the same perturbative input, namely the NLO ChPT results.
It is a matter of properly reshuffling the chiral expansion in a way clearly motivated by unitarity and analyticity.
We also show $|F_S(s)|^2$ in the right top panel of Fig.~\ref{fig.200220.1}, in which the resonance shape
due to the $f_0(500)$ is clearly visible. These resonance effects are not so evident in the case of the
isoscalar scalar phase shifts because of the Adler zero in this $\pi\pi$ PWA, which interferes strongly with the
pole contribution from the resonance itself.

\subsection{KT formalism}
\label{sec.200522.1}

The KT formalism was originally developed by Ref.~\cite{khuri} to study the $K\to 3\pi$ decays and,
up to including two-body intermediate states, it allows to implement unitarity and crossing symmetry.
Later on, this approach has been applied to study extensively the $\eta\to 3\pi$ decays, among others.
These decays violate isospin because the $G$ parity of the $\eta$ is $+1$ and that of the pion in $-1$,
so that it is  proportional to $m_u-m_d$ in pure QCD.

The application of ChPT to the $\eta\to 3\pi$ decays has been controversial, until accepting 
that FSI are so strong that a non-perturbative unitarization method is needed to be
implemented in order to be able to confront well with experimental data \cite{truong.200211.1}. 
The earliest calculations using current-algebra techniques obtained a value for the $\eta\to\pi^+\pi^-\pi^0$ of 
around 65~eV \cite{weinberg.190329.1}, too small compared with the experimental result
$\Gamma(\eta\to \pi^+\pi^+\pi^0)=(300 \pm 12)$~eV \cite{pdg}.
Roiesnel and Truong \cite{truong.200211.1} 
stressed that  a non-perturbative calculation taking care of the isoscalar-scalar
$\pi\pi$ FSI, by employing an Omn\`es function on top of the current-algebra result,
increases the decay width up to 200~eV. 
A few years later, the NLO ChPT calculation \cite{CT.181127.3} gives $(160\pm 50)~$eV,
which implies a large correction by a
factor 2.4 over the LO calculation in the right direction, but still too small by around  a factor of 2. 
In addition, the parameter $ \alpha$, typically employed in the parameterization of the Dalitz plot for the decay 
$\eta\to 3\pi^0$, is positive at NLO ChPT \cite{CT.181127.3} 
while experimentally it is negative, $\alpha=-0.0318\pm 0.0015$ \cite{pdg}. 
The calculation at NNLO in ChPT of the $\eta\to 3\pi$ decays  was performed in Ref.~\cite{bijnens.190329.1} but
the proliferation of new counterterms prevented a sharp result. 
If resonance saturation is assumed to estimate the NNLO ChPT counterterms then the Dalitz plot parameters
are not well reproduced.
One then concludes that the $\eta\to 3\pi$ decays are sensitive to the detailed values 
of the ${\cal O}(p^6)$ counterterms, so that an accurate calculation requires a precise knowledge of their values.  
This controversial situation stimulated  the interest in developing sophisticated calculations
combining ChPT and non-perturbative methods, within
unitarized ChPT \cite{truong.200211.1,borasoy.190329.1,borasoy.190329.2} and the KT formalism
\cite{kambor.190329.1,anisovich.190329.1,jpac.190329.1,colangelo.190329.1,alba.190329.1}.

We now describe the basic points of the one-channel KT formalism for $\eta\to 3\pi$ decays and refer the reader
to Refs.~\cite{descotes.190329.1,alba.190329.1} and the recent review \cite{oller.review} for further details.
In particular, the generalization to coupled channels was worked out in Ref.~\cite{alba.190329.1}, given in a
more compact matrix notation in Ref.~\cite{oller.review}.

Let us consider the decay $\eta(p_0)\to\pi^+(p_1)\pi^-(p_2)\pi^0(p_3)$, which is
related by crossing symmetry to the scattering reactions  $\eta(p_0)\pi^0(-p_3)\to\pi^+(p_1)\pi^-(p_ 2)$ in the $s$-channel,
$\eta(p_0)\pi^-(-p_1)\to\pi^0(p_3)\pi^-(p_ 2)$ in the $t$-channel,
and $\eta(p_0)\pi^+(-p_2)\to\pi^+(p_2)\pi^0(p_ 3)$ in the $u$-channel.
The Mandelstam variables $s$, $t$ and $u$ are given by
\begin{align}
\label{200522.1}
s&=(p_0-p_3)^2=(p_1+p_2)^2~,\\
t&=(p_0-p_1)^2=(p_2+p_3)^2~,\nn\\
u&=(p_0-p_2)^2=(p_1+p-3)^2~.\nn
\end{align}
The crossing-symmetry relations are 
\begin{align}
\label{200522.2}
T(\eta\pi^0\to \pi^+\pi^-)&=A(s,t,u)~,\\
T(\eta \pi^- \to \pi^0\pi^-)&=A(t,s,u)~,\nn\\
T(\eta \pi^+\to \pi^+\pi^0)&=A(u,t,s)~.\nn
\end{align}
These amplitudes in turn can be decomposed in scattering amplitudes with well defined isospin, $M^I(s,t,u)$, as
\begin{align}
\label{200522.3}
A(\eta\pi^0\to \pi^+\pi^-)=A(s,t,u)&=-\frac{1}{3}M^2(s,t,u)+\frac{1}{3}M^0(s,t,u)~,\\
A(\eta\pi^+\to \pi^+\pi^0)=A(u,t,s)&=+\frac{1}{2}M^2(s,t,u)+\frac{1}{2}M^1(s,t,u)~,\nn\\
A(\eta\pi^-\to \pi^0\pi^-)=A(t,s,u)&=+\frac{1}{2}M^2(s,t,u)-\frac{1}{2}M^1(s,t,u)~.\nn
\end{align}
The inversion of these relations gives us the $M^I(s,t,u)$,  
\begin{align}
\label{200522.4}
M^0(s,t,u)&=3A(s,t,u)+A(u,t,s)+A(t,s,u)~,\\
M^1(s,t,u)&=A(u,t,s)-A(t,s,u)~,\nn\\
M^2(s,t,u)&=A(u,t,s)+A(t,s,u)~.\nn
\end{align}
The PWA amplitudes are denoted by $M^{(IJ)}(s)$, and one has the standard relations
\begin{align}
\label{200522.5c}
M^{(IJ)}(s)&=\frac{1}{2}\int_{-1}^{+1}d\cos\theta\,P_J(\cos\theta)M^I(s,t,u)~,\\
M^I(s,t,u)&=\sum_{J=0}^\infty (2J+1)P_J(\cos\theta)M^{(IJ)}(\cos\theta)~.\nn
\end{align}
In the KT formalism the $S$ and $P$ waves are the ones that are subject to a non-perturbative
treatment.

The PWAs have a RHC above the two-pion threshold $s>4m_\pi^2$.
Instead of writing the unitarity constraint as in Eq.~\eqref{200304.3c}, one should consider it as giving
the discontinuity along the RHC because of the two on-shell intermediate pions. We then write\footnote{Due 
to the fact that in the decay channel all the three pions are on-shell in the region $(m_\eta-m_\pi)^2\geq s \geq 4m_\pi^2$ 
this is another source of imaginary part from the crossed-channel cuts that are also on-shell.
For $s=(m_\eta^2-m_\pi^2)/2$ the branch point singularity at $t,u=4m_\pi^2$
happens for $\cos\theta=\mp 1$. These crossed-channel cuts can be separated from the RHC one by 
giving a vanishing positive imaginary part to $m_\eta^2$ and then proceed by analytical continuation in
$m_\eta^2$ \cite{mandelstam.190403.1}.}
\begin{align}
\Im M^{(IJ)}&\to \frac{M^{(IJ)}(s+i\epsilon)-M^{(IJ)}(s-i\epsilon)}{2i}=\rho M^{(IJ)}(s+i\ep)M^{(IJ)}(s-i\ep)~,\nn \\
\label{200522.5a}
 M^{(IJ)}(s+i\ep)&=\underbrace{\left(1+2i\rho M(s+i\ep)\right)}_{{\text{$S$-matrix~in~PWAs}}}M(s-i\ep)=e^{2i\delta^{(IJ)}(s)}M(s-i\ep)~.
\end{align}
From this last line we can write more conveniently the discontinuity of $M^{(IJ)}(s)$ along the RHC, $\Delta M^{(IJ)}(s)$, as
\begin{align}
\label{200522.5}
\Delta M^{(IJ)}(s)&=M^{(IJ)}(s+i\ep)-M^{(IJ)}(s-i\ep)=2i\sin\delta^{(IJ)}e^{-i\delta^{(IJ)}}M^{(IJ)}(s+i\ep)~,
\end{align}
which is the relation finally used.

A crucial feature of the KT formalism is to write down $A(s,t,u)$ as the sum of
three functions of only one Mandelstam variable, $M_0(s)$, $M_1(s)$ and $M_2(t)$ \cite{alba.190329.1,colangelo.190329.1}
\begin{align}
\label{200522.6}
A(s,t,u)&=M_0(s)-\frac{2}{3}M_2(s)+(s-u)M_1(t)+(s-t)M_1(u)+M_2(t)+M_2(u)~,
\end{align}
which is invariant under the exchange $t\leftrightarrow u$, a feature that can be seen as a consequence of
charge-conjugate invariance. This representation is valid up to ${\cal O}(p^8)$ in ChPT
\cite{bijnens.190329.1,colangelo.190329.1} because then the $\pi\pi$ $D$ waves
also contribute and higher polynomials in $(s-t)$ and $(s-u)$ would be required.
The derivation of Eq.~\eqref{200522.6} can be understood by considering only $J\leq 1$ PWAs in the $s$-channel
and taking into account the isospin decomposition
for the process $\eta\pi^0\to \pi^+\pi^-$ and the crossed-channel ones, cf. Eq.~\eqref{200522.3}.
In this way, for the $s$-channel process there is no $I=1$ contribution, which only happens in the
crossed ones, cf. Eq.~\eqref{200522.3}. As this is a $P$-wave we then write it as $M_1(t)(s-u)+M_1(u)(s-t)$, that
also keeps explicitly the symmetry under the exchange $t\leftrightarrow u$. 
The $I=0$ contribution can only happen in the $s$-channel, because for the other channels the third component of isospin
is not zero.
This is the $M_0(s)$ contribution in Eq.~\eqref{200522.6}. Finally, regarding the $I=2$
it is clear from Eq.~\eqref{200522.3} that it appears in the combination
$-2 M_2(s)/3+M_2(t)+M_2(u)$.

Taking the expression for $A(s,t,u)$ in the ones of $M^{I}(s,t,u)$, as given in Eq.~\eqref{200522.4}, it follows that
\begin{align}
\label{200522.7}
M^0(s,t,u)&=3M_0(s)+M_0(t)+M_0(u)+\frac{10}{3}\big[M_2(t)+M_2(u)\big]+2(s-u)M_1(t)+2(s-t)M_1(u)~,\nn\\
M^1(s,t,u)&=2(u-t)M_1(s)+(u-s)M_1(t)-(t-s)M_1(u)+M_0(u)-M_0(t)+\frac{5}{3}\big[M_2(t)-M_2(u)\big]~,\nn\\
M^2(s,t,u)&=2M_2(s)+\frac{1}{3}\big[M_2(t)+M_2(u)\big]+M_0(t)+M_0(u)-(s-u)M_1(t)-(s-t)M_1(u)~.
\end{align}

Writing down the PWAs for $IJ=00,20$ and 11 we have
 \begin{align}
\label{200522.8}
M^{00}(s)&\equiv 3\big[M_0(s)+\hat{M}_0(s)\big]\\
M^{11}(s)&\equiv -\frac{2}{3}\kappa\big[ M_1(s)+\hat{M}_1(s)\big]~, \nn\\
M^{20}(s)&\equiv 2\big[M_2(s)+\hat{M}_2(s)\big]~,\nn
\end{align}
where
\begin{align}
\label{200522.9}
\kappa(s)=\sqrt{\sigma(s)\lambda(s)}
\end{align}
with
\begin{align}
\label{200522.10}
\lambda(s)&=\lambda(s,m_1^2,m_2^2)=s^2+m_\eta^4+m_\pi^4-2s(m_\pi^2+m_\eta^2)-2m_\pi^2m_\eta^2~,\\
\sigma(s)&=1-\frac{4m_\pi^2}{s}~.\nn
\end{align}
We have also introduced in Eq.~\eqref{200522.8} the angular averages
\begin{align}
\label{200522.11}
\langle M_I\rangle_n&=\frac{1}{2}\int_{-1}^{+1}d\cos\theta\, \cos\theta^n
M^I(s,t(s,\cos\theta),u(s,\cos\theta))~,\\ 
\hat{M}_0(s)&=
\frac{2}{3}\langle M_0\rangle_0+\frac{20}{9}\langle M_2\rangle_0+2(s-s_0)\langle M_1\rangle_0
+\frac{2}{3}\kappa \langle M_1\rangle_0~,\nn\\
\kappa(s)\hat{M}_1(s)&=
\frac{9}{2}(s-s_0)\langle M_1\rangle_1 +\frac{3\kappa}{2}\langle M_1\rangle_2
  +3\langle M_0\rangle_1-5\langle  M_2\rangle_1 ~,\nn\\
\hat{M}_2(s)&=
  \frac{1}{3}\langle M_2\rangle_0+ \langle M_0\rangle_0-\frac{3}{2}(s-s_0)\langle M_1\rangle_0
-\frac{\kappa}{2}\langle M_1\rangle_1~,\nn
\end{align}
  and
\begin{align}
\label{200522.12}
3s_0&=m_\eta^2+3m_\pi^2~.
\end{align}
The function $\hat{M}_I(s)$ has no discontinuity across the RHC so that the
discontinuities of the PWAs $M^{(IJ)}(s)$ can be expressed as,
\begin{align}
\label{200522.13}
\Delta M_0(s)&=2i e^{-i\delta^{(00)}(s)}\sin\delta^{(00)}(s) \big[M_0(s)+\hat{M}_0(s)\big]~,\\
\Delta M_1(s)&=2i e^{-i\delta^{(11)}(s)}\sin\delta^{(11)}(s) \big[M_1(s)+\hat{M}_1(s)\big]~,\nn\\
\Delta M_2(s)&=2i e^{-i\delta^{(20)}(s)}\sin\delta^{(20)}(s) \big[M_2(s)+\hat{M}_2(s)\big]~.\nn
\end{align}

Following the same steps as above in  Eq.~\eqref{200522.5} we can then also write that
\begin{align}
\label{200522.14}
M_I(s+i\ep)&=M_I(s-i\ep)e^{2i \delta_{IJ}}+2i\hat{M}_I(s)e^{i\delta^{(IJ)}}sin\delta^{(IJ)}~,
\end{align}
with $J=I$ except for $I=2$ for which $J=0$ (as it should be clear from the context in this section).
Dividing this expression by the corresponding Omn\`es function $\Omega^{(IJ)}(s)$, which fulfills that along
the RHC $\Omega^{(IJ)}(s+i\ep)=e^{i2\delta^{(IJ)}}\Omega^{(IJ)}(s-i\ep)$, $\Omega^{(IJ)}(s+i\ep)=|\Omega^{(IJ)}(s)|e^{i\delta^{(IJ)}}$,
we then obtain from Eq.~\eqref{200522.14} the
discontinuity of $M_I/\Omega^{(JI)}$ as
\begin{align}
\label{200522.15b}
\frac{M_I(s+i\ep)}{\Omega^{(IJ)}(s+i\ep)}-\frac{M_I(s-i\ep)}{\Omega^{(IJ)}(s-i\ep)}&
=2i\frac{\hat{M}_I(s)\sin\delta^{(IJ)}}{|\Omega^{(IJ)}(s)|}~.
\end{align}
The final step is to obtain IEs for $M_I(s)$ by writing down DRs for $M_I(s)/\Omega^{(IJ)}$ as
\begin{align}
\label{200522.15}
M_I(s)&=\Omega^{(IJ)}(s)\Big[
P_I^{(m)}(s)+\frac{s^n}{\pi}\int_{4m_\pi^2}^{\infty}ds'\frac{\hat{M}_I(s') \sin\delta^{(IJ)}(s')}{|\Omega^{(IJ)}(s')|(s')^n(s'-s)}\Big]~,
\end{align}
where $P_I^{(m)}(s)$ is a subtractive polynomial with $m\geq n-1$. Requiring that $A(s,t,u)$ diverges linearly at most at infinity in the
Mandelstam variables \cite{anisovich.190329.1},
then $M_1(s)$ should be bounded by a constant and $M_0(s)$, $M_2(s)$ would diverge linearly at most
in the limit $s\to\infty$. Furthermore, we also know the asymptotic behavior in the same limit for the Omn\`es functions,
cf. Eq.~\eqref{181122.3}, with $|\Omega^{(IJ)}(s)|\to s^{-\delta^{(IJ)}(\infty)/\pi}$. Depending on $\delta^{(IJ)}(\infty)$
the value of $m$ should be adjusted to the required asymptotic behavior of $M_I(s)$. For instance, 
Ref.~\cite{alba.190329.1} assumes that $\delta^{(00)}(\infty)=\pi$, $\delta^{(11)}(\infty)=\pi$ and $\delta^{(20)}=0$, so that
$m=2$ for $I=0$ and $m=1$ for $I=1,$ and 2.

The DRs in Eq.~\eqref{200522.15} constitute a set of coupled linear IEs because the angular averages $\langle \hat{M}_I\ra_n$
are also expressed in terms of the $M_I(s)$ functions. A standard way for solving these equations is by iteration.
The subtraction constants can be determined by matching  with the NLO ChPT calculation of $A(s,t,u)$ and/or fitted to data,
as done in Refs.~\cite{anisovich.190329.1,alba.190329.1}. A clear improvement is obtained in the calculated decay width for the
$\eta\to\pi^+\pi^-\pi^0$ in Ref.~\cite{anisovich.190329.1}, where the value $\Gamma_{\eta\to\pi^+\pi}=283\pm 28$~eV was obtained.
Other improvements concern the parameter $\alpha$ for characterizing the amplitude for $\eta\to 3\pi^0$ in its Dalitz plot.
NLO ChPT gives a value $\alpha=0.0142$ while the KT treatment of Ref.~\cite{alba.190329.1}
gives $\alpha=-0.0337(12)$, to be compared with 
the PDG average value of $\alpha=-0.0318(15)$.

\section{The $N/D$ method}
\setcounter{equation}{0}   
\label{sec.200517.1}

In this section we elaborate on different aspects of the $N/D$ method, first introduced in Ref.~\cite{chew} to study uncoupled $\pi\pi$ PWAs. We first review on this method, discuss in more detail  the limit
in which the crossed-channel dynamics is neglected \cite{nd}, and afterwards elaborate on how the latter can be treated
perturbatively within the $N/D$ method \cite{plb,ww}.
These results can also be used to take into account FSI in production
processes \cite{palomar,ozi}.  
For the case of NR scattering, thanks to recent developments
\cite{oller.lhc.aop}, it is possible to know the
exact discontinuity of a PWA along the LHC for a given potential. In this way, one can
generate the same solutions as in the Lippmann-Schwinger (LS) equation, together with other ones
that cannot be obtained in a LS equation when mimicking the short-distance interactions
by contact terms in the potential \cite{oller.lhc.plb}.

\subsection{Scattering}
\label{sec.200517.2}

For the scattering of particles with equal masses  there is only a 
LHC for $s<s_{\rm Left}$ because of crossing. 
However, when the particles involved have different masses there are also
other types of cuts in the complex $s$ plane due to crossing. For instance,
for the scattering of particles $a+b\rightarrow a+b$,  in addition to a LHC there is also
a circular cut  for $|s|=m_b^2-m_a^2$ \cite{martin.290916.1} where, for definiteness,
we have considered that $m_b>m_a$. 
Nonetheless, when we refer in the following  to the LHC we actually mean all the crossed-channel cuts.
Indeed, had we taken instead the  complex $p^2$ plane all the cuts would be linear and only
a LHC would be present \cite{martin.290916.1}.

We introduce the $N/D$ method following Ref.~\cite{nd}. 
The uncoupled case is discussed first and afterwards we move to coupled-channel scattering.
The discussion is restricted to two-body intermediate states. 
The discontinuity of the inverse of a PWA $T_\ell(s)$ along the RHC is $2i$ times its imaginary part,  
being the latter fixed by phase space because of unitarity, cf. Eq.~\eqref{200304.3b}.

In the $N/D$ method   $T_\ell(s)$  is expressed as the quotient of two functions,
\begin{align}
\label{n/d}
T_\ell(s)=\frac{N_\ell(s)}{D_\ell(s)}~,
\end{align}
where $N_\ell(s)$ stands for the numerator function and $D_\ell(s)$ for the denominator one.
The former has only LHC and the later RHC. 
 
To enforce the right kinematical threshold behavior of a PWA, vanishing as $p^{2\ell}$,  
Ref.~\cite{nd} divides $T_\ell(s)$ by $p^{2\ell}$,  
\begin{align}
\label{T'}
T'_\ell(s)=\frac{T_\ell(s)}{p^{2\ell}}~.
\end{align} 
The $N/D $ method is then applied to this function, 
\begin{align}
\label{n/d'}
T'_\ell(s)=\frac{N'_\ell(s)}{{D}'_\ell(s)}~.
\end{align}
It follows then  that the discontinuities of ${N}'_\ell(s)$ and 
${D}'_\ell(s)$ along the LHC and RHC, respectively, are
\begin{align}
\label{eqs1} 
{\Im  D}'_\ell &=\Im T'^{-1}_\ell \; {N}'_\ell=-\rho(s) {N}'_\ell p^{2\ell}~,  &s>s_{\rm th}~, \\
{\Im D}'_\ell&=0~,   &s<s_{\rm th}~,\nn\\
\label{eqs2}
{\Im  N}'_\ell&=\Im T'_\ell \; {D}'_\ell=\Delta_\ell D'_\ell {p^{2\ell}}~,  &s<s_{\rm Left} ~,  \\
{\Im N}'_\ell&=0~,  &s>s_{\rm Left}~,\nn
\end{align}
with $\Delta_\ell(s)=\Im T_\ell(s)$ along the LHC.
Let us discuss the DRs for  ${D}'_\ell(s)$ and ${N}'_\ell(s)$ that result by taking into account
these discontinuities. 
For $D'_\ell(s)$ one has,
\begin{align}
\label{d'}
{D}'_\ell(s)=\sum_{m=0}^{n-1}\overline a _m s^m-\frac{(s-s_0)^n}{\pi}\int^\infty_{s_{\rm th}} ds' 
\frac{p(s')^{2\ell} \rho(s') {N}'_\ell(s')}{(s'-s)(s'-s_0)^n}~.
\end{align}
Here $n$ is, at least, the minimum  number of subtractions required to guarantee the convergence of the
integral in the DR, 
\begin{align}
\label{n}
\displaystyle \lim_{s \to \infty} \frac{{N}'_\ell(s)}{s^{n-\ell}}= 0~.
\end{align}

Consistently with  Eq.~(\ref{n}), the DR for $N'_\ell(s)$ can be written as                                      
\begin{align}
\label{n2}
{N}'_\ell(s)=\sum_{m=0}^{n-\ell-1} \overline b_m s^m+\frac{(s-s_0)^{n-\ell}}{\pi}\int_{-\infty}^{s_{\rm Left}} ds' 
\frac{\Delta_\ell(s') {D}'_\ell(s')}{p(s')^{2\ell} (s'-s_0)^{n-\ell} (s'-s)}~.
\end{align}                         
The Eqs.~(\ref{d'}) and (\ref{n2}) are a system of coupled linear IEs whose input is $\Delta_\ell(s)$.
It is customary to substitute the expression for $N'_\ell(s)$ in $D'_\ell(s)$ and end with a linear IE for 
$D'_\ell(s)$ along the LHC. Namely,
\begin{align}
\label{200519.5}
D'_\ell(s)&=\sum_{m=0}^{n-1}\overline{a}_m s^m
-\sum_{m=0}^{n-\ell-1}\overline b_m\frac{(s-s_0)^n}{\pi}\int_{s_{\rm th}}^\infty ds'\frac{p(s')^{2\ell}\rho(s'){s'}^m}{(s'-s)(s'-s_0)^n}\\
&+\frac{(s-s_0)^n}{\pi^2}\int_{-\infty}^{s_{\rm Left}}ds''\frac{\Delta_\ell(s'')D_\ell'(s'')}{(s''-s_0)^{n-\ell}p(s'')^{2\ell}}
\int_{s_{\rm th}}^\infty ds'\frac{p(s')^{2\ell}\rho(s')}{(s'-s)(s'-s'')(s'-s_0)^\ell }~,\nn
\end{align}
and the last integral can indeed be performed algebraically.
This is a linear IE for $D'_\ell(s)$ with $s$ along the LHC. Once this solved one can calculate
$D_\ell(s)$ for $s\in \mathbb{C}$ and, in particular, along the physical region, $s+i\ep$.
Other types of IEs could be deduced by taking more subtractions independently in $D_\ell(s)$ and $N_\ell(s)$.
Fore more details on this respect the reader can consult \cite{guo.rios.NN}.

The expression in Eq.~\eqref{200519.5} can be shortened and simplified  for equal mass scattering
with mass $m$ by taking $s_0=4m^2$, 
because then $p(s')^2=(s-4m^2)/4$. It follows that, 
\begin{align}
\label{200519.6}
D'_\ell(s)&=\sum_{m=0}^{n-1}\overline{a}_m s^m
-\sum_{m=0}^{n-\ell-1}\overline b_m\frac{(s-s_0)^n}{4^\ell \pi}\int_{s_{\rm th}}^\infty ds'\frac{\rho(s'){s'}^m}{(s'-s)(s'-s_0)^{n-\ell}}\\
&+\frac{(s-s_0)^n}{\pi^2}\int_{-\infty}^{s_{\rm Left}}ds''\frac{\Delta_\ell(s'')D_\ell'(s'')}{(s''-s_0)^{n}}
\int_{s_{\rm th}}^\infty ds'\frac{\rho(s')}{(s'-s) (s'-s'')}~.\nn
\end{align}
The last integral in the previous expression can be written in terms of $g(s)$, Eq.~\eqref{200508.2}.

One of the subtraction constants can be fixed because we can freely choose the normalization of $D'_\ell(s)$, since
their ratio and analytical properties are invariant under a change in normalization. The standard choice is to take $D'_\ell(0)=1$. 
However, given $\Delta'_\ell(s)$ along the LHC, the solution is not unique because of the addition of extra subtraction constants
in $D'_\ell(s)$ and $N'_\ell(s)$.

Historically, the possible addition of Castillejo-Dalitz-Dyson (CDD) poles \cite{Castillejo} was the clear
indication that extra solutions could be obtained even if  $\Delta_\ell(s)$ is assumed to be known along the LHC. 
They give rise to zeros of $T_\ell(s)$ along the RHC and each zero comprises two real parameters, its residue and 
position. Phenomenologically the CDD poles
correspond to the short-distance dynamics  underneath the scattering process and
might also be related to the addition of bare states \cite{dyson}.
Let us notice that $T_\ell(s)^{-1}$ does not exist at a zero of $T_\ell(s)$ and, therefore,  Eq.~\eqref{200304.3b}
is not defined there. As in Ref.~\cite{Castillejo} let us introduce the auxiliary function $\lambda(s)$ such that 
\begin{align}
\label{dl1}
{\Im D}'_\ell(s)=\frac{d \lambda (s)}{ds}~,
\end{align} 
and rewrite Eq.~(\ref{eqs1}) as, 
\begin{eqnarray}
\label{dl2}
{\displaystyle \frac{d\lambda}{ds}}&=-\rho(s) p^{2\ell}{N}'_\ell~, & s>s_{\rm th}~, \\
{\displaystyle \frac{d\lambda}{ds}}&=0~.~~~~~~~~~~~~~~ & s<s_{\rm th}~. \nn
\end{eqnarray}

Denoting by  $s_i$ the zeros of $T_\ell(s)$ along the real axis above threshold, 
we can write $\lambda(s)$ from Eq.~\eqref{dl2} as  
\begin{align}
\label{l2}
\lambda(s)=-\int_{s_{\rm th}}^s p(s')^{2\ell} \rho(s') {N}'_\ell(s') ds'+
\sum_i \lambda(s_i)\theta(s-s_i)~,
\end{align}
where the $\lambda(s_i)$ are a priori unknown.
Thus, Eqs.~(\ref{dl1}) and (\ref{l2}) allow us to write 
\begin{align}
\label{tower1}
{D}'_\ell(s) &= \sum_{m=0}^{n-1} \overline a_m s^m + 
\frac{(s-s_0)^n}{\pi}\int_{s_{\rm th}}^\infty \frac{
	{\Im  D}'_\ell(s') ds'}{(s'-s)(s'-s_0)^n} \\ 
& =\sum_{m=0}^{n-1} \overline a_m
s^m -\frac{(s-s_0)^n}{\pi}\int_{s_{\rm th}}^\infty \frac{p(s')^{2\ell} \rho(s') 
	{N}'_\ell(s')}{(s'-s)(s'-s_0)^n}ds' +
\frac{(s-s_0)^n}{\pi} \int_{s_{\rm th}}^\infty \frac{\sum_i \lambda(s_i) 
	\delta(s'-s_i)}{(s'-s)(s'-s_0)^n}ds'\nonumber\\
&=\sum_{m=0}^{n-1} \overline a_m s^m
-\frac{(s-s_0)^n}{\pi}\int_{s_{\rm th}}^\infty \frac{p(s')^{2\ell} \rho(s') 
	{N}'_\ell(s')}{(s'-s)(s'-s_0)^n}ds'+\sum_i \frac{\lambda(s_i)}{\pi(s_i-s) }\frac{(s-s_0)^n}{(s_i-s_0)^n}~.\nonumber
\end{align}

The last term in the previous equation can be rewritten as
\begin{align}
\label{cano}
\frac{(s-s_0)^n}{s-s_i}&=\sum_{i=0}^{n-1}(s-s_0)^{n-1-i}(s_i-s_0)^i+\frac{(s_i-s_0)^n}{s-s_i}~.
\end{align}
The contribution $\displaystyle{\sum_{i=0}^{n-1}(s-s_0)^{n-1-i}(s_i-s_0)^i}$ can  be reabsorbed in  
$\displaystyle{\sum_{m=0}^{n-1} \overline a _m s^m}$ and Eq.~\eqref{tower1} can be rewritten as  
\begin{align}
\label{d'2}
{D}'_\ell(s)&=\sum_{m=0}^{n-1} \widetilde{a}_m s^m + 
\sum_i \frac{\widetilde{\gamma}_i}{s-s_i}
-\frac{(s-s_0)^n}{\pi}\int_{s_{\rm th}}^\infty 
\frac{p(s')^{2\ell} \rho(s') {N}'_\ell(s')}{(s'-s)(s'-s_0)^n}ds' ~,
\end{align} 
where $\widetilde{a}_m$,  $\widetilde{\gamma}_i$ and $s_i$  are constants not fixed by the knowledge of $\Delta_\ell(s)$,
and the CDD poles give rise to the last term.  

 Interesting results can be deduced under the approximation of neglecting the LHC,
$\Delta_\ell(s)\to 0$. Eq.~(\ref{n2}) then becomes
\begin{align}
\label{naprox}
{N}'_\ell(s)=\sum_{m=0}^{n-\ell-1} \overline{b}_m s^m=\overline b_{n-\ell-1} \prod_{j=1}^{n-\ell-1}(s-s_j)~,
\end{align} 
and $N'_\ell(s)$ is just a polynomial, which can be reabsorbed in $D'_\ell(s)$ by dividing simultaneously both functions by
$N'_\ell(s)$ itself. The expression for $T'_\ell(s)$ then becomes
\begin{eqnarray}
\label{fin/d}
{T}'_\ell(s)&=&\frac{1}{{D}'_\ell(s)}~,\\
{N}'_\ell(s)&=&1~,\nn\\
{D}'_\ell(s)&=&-\frac{(s-s_0)^{L+1}}{\pi}\int_{s_{\rm th}}^\infty  
\frac{p(s')^{2\ell} \rho (s')}{(s'-s)(s'-s_0)^{L+1}}ds'+\sum_{m=0}^L a_m s^m+
\sum_i^{M_\ell} \frac{R_i}{s-s_i}~.\nn
\end{eqnarray}
The number of real free parameters in the previous equation  is $\ell+1+2M_\ell$, with $M_\ell$ the number of
CDD poles.  A priori there is nothing to prevent the generalization of Eq.~\eqref{fin/d} such that
some $s_i$ could also lie below threshold.  
We could adjust the position and residue of a CDD pole such that the real part of
$D'_\ell(s)$ vanishes  at the desired position.
This would give rise to typical resonance behavior above threshold, or to a bound-state pole if
this happens below threshold.   
This is why the parameters of the CDD poles are typically associated with the coupling constants and masses 
of the poles in the $S$ matrix.  
In other instances,  the CDD poles are needed because the presence of a zero cannot be related
to $\Delta_\ell(s)$, but they respond to fundamental constraints in the theory.
This is the case of the Adler zeroes in QCD \cite{adler.181115.1}, which already
occur at LO in the chiral expansion,
while $\Delta_\ell(s)\neq 0$ only at NLO and higher orders.
It is therefore necessary to account for them by including CDD poles,
such that the derivative of the PWA at the zero corresponds to 
the inverse of the residue of the CDD pole, $R_i$. 
For the $\pi\pi$ Adler zeroes the latter could be fixed in good approximation by the LO ChPT result. 
The other $\ell+1$ parameters emerge by having enforced the correct behavior of a PWA near threshold,
which should vanish as $p^{2\ell}$. 

Let us stress that Eq.~(\ref{fin/d})  gives the general form of an elastic PWA
when the LHC contributions are neglected. 
Phenomenologically this assumption could be suited if the LHC is far away and/or if
it is suppressed for some reason \cite{X3872}. 
The free parameters in Eq.~\eqref{fin/d} can be fixed by fitting experimental data
and/or by reproducing the Lattice QCD (LQCD) results at finite
volume or when varying some of the QCD parameters, like $N_c$ or the quark masses
\cite{guo.190126.1,guo.190126.3,albaladejo.190126.1,roca.old}. 

Ref.~\cite{nd}  focuses on meson-meson scattering, whose basic theory is QCD.
It studied the $S$- and $P$-wave two-body scattering between the lightest pseudoscalars ($\pi$, $K$ and $\eta$),  
as well as the related spectroscopy. 
It was found that the full nonet of scalar resonances \cite{oller.mix} $f_0(500)$, $f_0(980)$, $a_0(980)$ and $K^*_0(800)$ arose
from the self-interactions among the lightest pseudoscalars, while the more massive
resonances $f_0(1370)$, $f_0(1500)$, $a_0(1450)$ and $K^*_0(1430)$ stem from a nonet of bare resonances
with a mass around 1.4~GeV.
In addition, Ref.~\cite{nd} included a bare scalar singlet with a mass around 1~GeV which
gives also a contribution to the $f_0(980)$  \cite{guo.nc.traj}. 
Later on, Ref.~\cite{alba.oller.glue} extended this model by including more channels
and could determine a glueball state
affecting mainly the $f_0(1700)$ with a reflection (because of the $\eta\eta'$ threshold)
on the $f_0(1500)$ as well.  
Of course, the same Eq.~\eqref{fin/d} can be applied to other interactions,
e.g. Ref.~\cite{ww} studied $W_LW_L$ scattering in the electroweak symmetry breaking sector.

The generalization of Eq.~\eqref{fin/d} to coupled channels is rather straightforward by employing a
matrix notation, where 
the $T$ matrix in coupled channels is a matrix denoted by $T_L(s)$.
As in Eq.~\eqref{fin/d} we take from the onset that crossed-channel dynamics can be
neglected in a first approximation.
Thus, the matrix element ${T}_{L,ij}(s)$ is proportional to $p_i^{\ell_i} p_j^{\ell_j}$,
which gives rise for odd orbital angular momentum (unless $i=j$) 
to another cut between $s_{{\rm th}; i}$ and $s_{{\rm th}; j}$
due to the square roots in the expressions of $p_i$ and $p_j$ as a
function of $s$.
To avoid this cut we define the matrix $T'_L$, analogously to Eq.~\eqref{T'}, as 
\begin{align}
\label{A.1}
{T}'_L(s)=p^{-L} {T}_L(s) p^{-L}~.
\end{align}
In this equation, the symbol $p^L$ corresponds to a diagonal matrix with matrix elements
\begin{align}
p^L_{ij}&=p_i^{\ell_i} \delta_{ij}~,\\ 
p_i&=\frac{\lambda^{1/2}(s,m_{1i}^2,m_{2i}^2)}{2\sqrt{s}}~,\nn 
\end{align}
and $m_{1i}$ and $m_{2i}$ are the masses of the two particles in the same channel $i$. 
The matrix unitarity relation along the RHC then reads 
\begin{align}
\label{A.2}
{\Im T}'^{ -1}_L(s)=-p^L \rho(s) p^L=-\rho(s) p^{2L}~,
\end{align} 
where $\rho(s)$ is another diagonal matrix whose elements are $\rho_i(s)$.
The next step proceeds with the generalization to coupled channel of Eq.~\eqref{n/d} by writing ${T}'_L$ as 
\begin{align}
\label{A.4}
{T}'_L={D}'^{ -1}_L {N}'_L~,
\end{align}
with  ${N}'_L$ and ${D}'_L$ two matrices, the former only involves LHC and the later RHC, respectively. 
In our present case without LHC, the matrix elements of $N'_L$ are polynomials functions. Multiplying  
$N'_L$ and $D'_L$ in Eq.~\eqref{A.4}  to the left by ${N'_L}^{-1}$ we can always make that ${N}'_L=I$ and write, 
\begin{eqnarray}
\label{A.8}
{T}'_L&=&\widetilde{{D}}'^{-1}_L~, \\
\widetilde{{N}}'_L&=&I~, \nonumber\\
\widetilde{{D}}_L'&=&-\frac{(s-s_0)^{L+1}}{\pi}\int_0^\infty ds'
\frac{\rho(s')p^{2L}(s')}{(s'-s)(s'-s_0)^{L+1}}+{R}(s)~,\nonumber
\end{eqnarray}
with ${R}(s)$ a matrix of rational functions which poles produce the CDD poles
in $\widetilde{{D}}_L'$.
Let us notice that all the zeros in ${\rm det}{T}'_L$ correspond to CDD poles in the ${\rm det}\widetilde{{D}}_L'(s)$.
This is the generalization  of the CDD poles for the coupled-channel case.

The resulting expression for $T_L(s)$ in Eq.~\eqref{A.8} can be also recast as
\begin{align}
\label{200518.1}
T_L(s)&=\left[V_L^{-1}+g(s)\right]^{-1}~,
\end{align}
with $g(s)$ the diagonal matrix with matrix elements $g_i(s)$ defined as
\begin{align}
\label{200518.2}
g_i(s)&=a_i(s_0)-\frac{s-s_0}{\pi}\int_{s_{{\rm th};i}}^\infty \frac{\rho_i(s')ds'}{(s'-s_0)(s'-s)}~,
\end{align}
where $a_i(s_0)$ is a subtraction constant and $s_0$ the subtraction point.
The result of this integration can also be written as
\begin{align}
\label{200518.3}
g_i(s)
&=\frac{1}{16\pi^2}\left[
a_i(\mu)+\log\frac{m_{1i}^2}{\mu^2}-x_+\log\frac{x_+-1}{x_+}-x_-\log\frac{x_--1}{x_-}\right]~,\nn\\
x_\pm&=\frac{s+m_{2i}^2-m_{1i}^2}{2s}\pm\frac{1}{2s}\sqrt{(s+m_{2i}^2-m_{1i}^2)^2-4s(m_{2i}^2-i0^+)}~.
\end{align}
The parameter $\mu$ is a renormalization scale, such that a change in the value of $\mu$ can always be 
reabsorbed in a corresponding variation of $a_i(\mu)$, while the combination $a_i(\mu)-2\log\mu$ is independent of $\mu$. 
The unitarity loop function $g_i(s)$ corresponds to the one-loop two-point function
\begin{align}
\label{181106.5}
g_i(s)&=i\int\frac{d^4p}{(2\pi)^4}\frac{1}{[(P/2-p)^2-m_{1i}^2+i\ve][(P/2+p)^2-m_{2i}^2+i\ve]}\\
&=\int_0^\infty \frac{p^2 dp}{(2\pi)^2}\frac{\omega_1+\omega_2}{\omega_1\omega_2[s-(\omega_1+\omega_2)^2+i\ve]}~,\nn
\end{align}
where $\omega_j=\sqrt{m_{ji}^2+\vp^2}$ and the total four-momentum $p_1+p_2$  is indicated by $P$. 
The integral in Eq.~\eqref{181106.5} diverges logarithmically, which is the reason why a subtraction
has been taken in Eq.~\eqref{200518.2}. 
The Eq.~\eqref{200518.3} also results by employing dimensional regularization  
and reabsorbing  the diverging term in $a_i(\mu)$.

Let us elaborate on the so-called natural value for the subtraction constants.
The function $g_i(s)$ given by Eq.~\eqref{200518.3} has the value  at threshold,  
\begin{align}
\label{200518.4}
g_i(s_{\rm th})&=\frac{a_i(\mu)}{16\pi^2}+\frac{1}{8\pi^2(m_{1i}+m_{2i})}(m_{1i}\log\frac{m_{1i}}{\mu}+m_{2i}\log\frac{m_{2i}}{\mu})~.
\end{align}
This expression is compared with the one that results by evaluating $g_i(s)$ in terms of a three-momentum cutoff  $\Lambda$.
The resulting expression for the function $g_i(s)$, and  denoted by $g_{\Lambda i}(s)$,
can be found in Ref.~\cite{iam.oller.long}.  
The natural size of a three-momentum cutoff in hadron physics  is the inverse of the typical size of a
compact hadron, which is generated by the strong dynamics binding quarks and gluons. 
Thus, according to this estimate we take  $\Lambda \simeq 1$~GeV.  
For NR scattering $g_i(s)$ and $g_{\Lambda i}(s)$ ($m_{1i},\,m_{2i} \gg |\vp|$)
are given by the value at threshold of every function plus
$-ip/(8\pi(m_1+m_2))+{\cal O}(\vp^2)$ \cite{guo.rios.NN}. 
The value at threshold of $g_{\Lambda i}(s_{\rm th})$ can be worked out 
explicitly  with the result \cite{guo.190126.1}
\begin{align}
\label{200518.5}
g_{\Lambda i}(s_{\rm th})&=-\frac{1}{8\pi^2(m_{1i}+m_{2i})}\left[
m_{1i}\log\left(1+\sqrt{1+m_{1i}^2/\Lambda^2}\right)\right.\\
&\left. +m_{2i}\log\left(1+\sqrt{1+m_{2i}^2/\Lambda^2}\right)
-m_{1i}\log\frac{m_{1i}}{\Lambda}-m_{2i}\log\frac{m_{2i}}{\Lambda}
\right]~.\nn
\end{align}     
By equating Eqs.~\eqref{200518.4} and \eqref{200518.5} the following matching value for $a_i(\mu)$ results, 
\begin{align}
\label{181106.8}
a_i(\mu)&=-\frac{2}{m_{1i}+m_{2i}}\left[m_{1i}\log\left(1+\sqrt{1+m_{1i}^2/\Lambda^2}\right)
+m_{2i}\log\left(1+\sqrt{1+m_{2i}^2/\Lambda^2}\right)    \right]+\log\frac{\mu^2}{\Lambda^2}~. 
\end{align}
One should employ $\mu\simeq \Lambda\simeq 1$~GeV in Eq.~\eqref{181106.8} to estimate the natural value for
the subtraction constants, a procedure originally established in Ref.~\cite{plb}. 
In this way, both the renormalization scale $\mu$ and the cut off $\Lambda$ are used with values
suitable to the transition from the low-energy EFT 
to the shorter-range QCD degrees of freedom.
As an example, let us take $\pi\pi$ scattering and $\Lambda=1$~GeV. Then, from Eq.~\eqref{181106.8}
\begin{align}
\label{181106.9}
a(\mu)&=-1.40+\log\frac{\mu^2}{\Lambda^2}~,~\Lambda=1~{\rm GeV}~.
\end{align}

The Eq.~\eqref{200518.1} is adequate for including perturbatively the LHC contributions in the $T$-matrix $T_L$.
This can be achieved by matching order by order with a calculation within an EFT. For instance, this has been used 
many times taking as input one-loop calculations in ChPT
\cite{ww,plb,jamin,guo.181123.1,alba.oller.sigma,guo.prc,guo.190126.1,guo.190126.3,talks,kangP}.
The procedure is as follows.
Let us take a meson-meson scattering amplitude calculated in ChPT up to one-loop or ${\cal O}(p^4)$,
$T_L= T_2+T_4+{\cal O}(p^6)$. Then the chiral expansion of Eq.~\eqref{200518.1}, with $V=V_2+V_4+{\cal O}(p^6)$, $g={\cal O}(p^0)$ \cite{ww},
reads at LO,
\begin{align}
\label{200519.1}
T_2&=V_2+{\cal O}(p^4)~,
\end{align}
and at NLO,
\begin{align}
\label{200519.2}
T_4=V_4-V_2 g V_2+{\cal O}(p^6)~,
\end{align}
and similarly for higher orders. 
Thus, up to NLO the matching equations fix $V_2$ and $V_4$ to 

\begin{align}
\label{200519.3}
V_2&=T_2~,\\
V_4&=T_4+V_2 gV_2~.\nn
\end{align}
The LHC contributions arise because crossed-channel loops are calculated order by order in the ChPT
results for $T_L$. 
At NLO in the calculation of $V_L$ we then have the expression
\begin{align}
\label{200613.1}
T_L(s)&=\left[(T_2+T_4+T_2 g T_2)^{-1}+g\right]^{-1}~.
\end{align}
If $(T_2+T_4+T_2 g T_2)^{-1}$ is further expanded we then recover the IAM result of Eq.~\eqref{200515.5} because
\begin{align}
\label{200613.2}
(T_2+T_4+T_2 g T_2)^{-1}+g=T_2^{-1}-T_2^{-1}T_4T_2^{-1}+{\cal O}(p^2)~,
\end{align}
so that 
\begin{align}
\label{200613.3}
T_L(s)\to \left[T_2^{-1}-T_2^{-1}T_4T_2^{-1}\right]^{-1}=T_2^{-1}\left[T_2-T_4\right]^{-1} T_2~.
\end{align}
This is the formula for the IAM in coupled channels at NLO \cite{iam.oller.long,oop.prl}.

\begin{figure}[H]
\begin{center}
\includegraphics[width=0.85\textwidth]{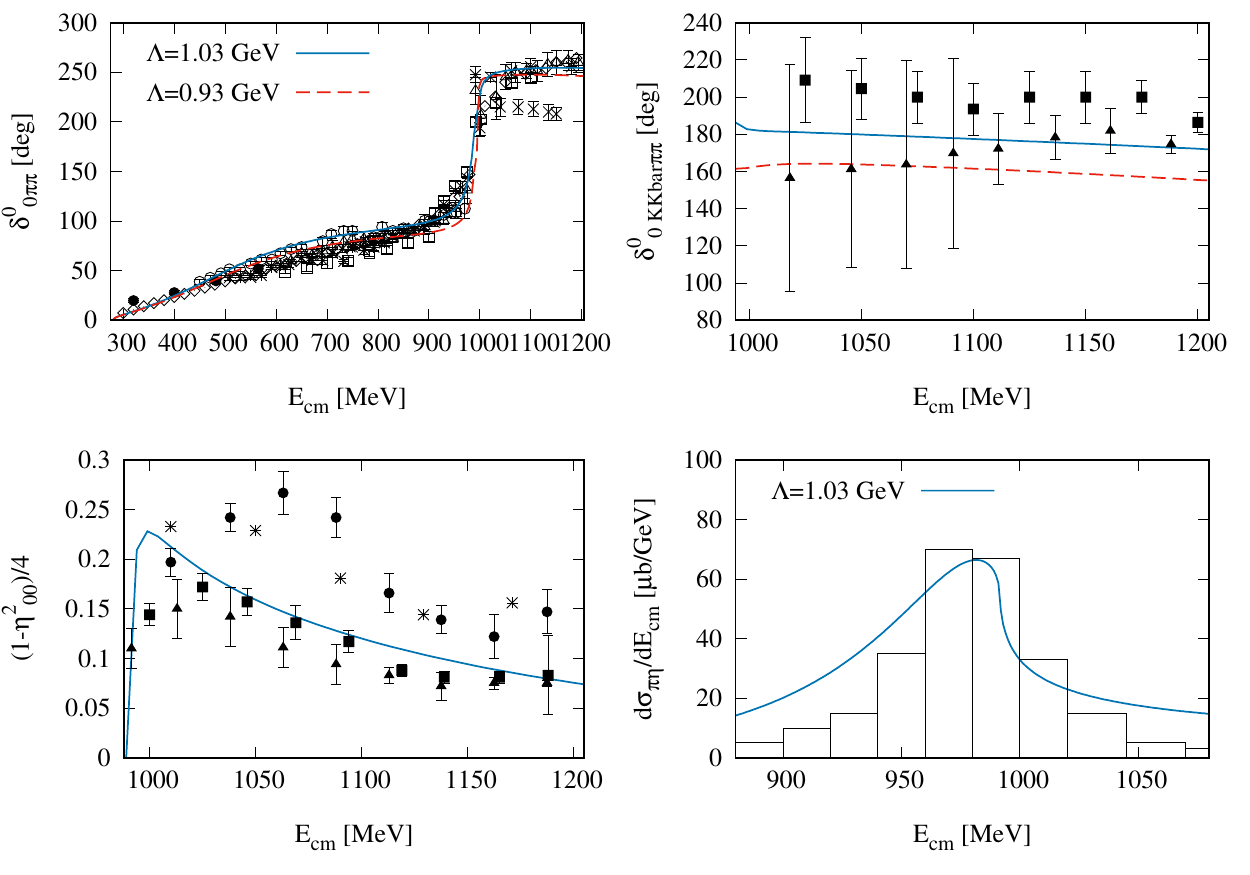} 
\caption{{\small  Results from Ref.~\cite{npa} with only one free parameter
    for the $S$-wave meson-meson scattering with $I=0$ and 1.
From top to bottom and left to right, the isoscalar scalar $\pi\pi\to \pi\pi$ and $K\bar{K}\to \pi\pi$ 
phase shifts, the $\pi\pi$ inelastic cross-section with the same quantum numbers
and a $\pi^0\eta$ event distribution around the isovector scalar
$a_0(980)$ resonance are plotted. 
For more details and references of the experimental papers we refer to Ref.~\cite{npa}.}
\label{fig.200524.1}}
\end{center}
\end{figure}

In order to appreciate the power of the method for some reactions we consider the LO matching, that is,
with $V=V_2$, applied in Ref.~\cite{npa} to study the meson-meson $S$-waves with $I=0$ and 1.
This is a coupled-channel study  with $\pi\pi$ and $K\bar{K}$ for $I=0$ and $\pi\eta$ and $K\bar{K}$ for $I=1$.
It is certainly remarkable that only one free parameter entered in the successful calculation of the PWAs 
from the $\pi\pi$ threshold up to around 1.2~GeV. This is shown in Fig.~\ref{fig.200524.1}
by the $\pi\pi\to \pi\pi$, $K\bar{K}\to\pi\pi$ phase shifts,
the inelastic $\pi\pi$ reaction and a $\pi^0\eta$ event distribution around the $a_0(980)$, from
top to bottom and left to right, respectively. 
The resonances $f_0(500)$, $f_0(980)$ and $a_0(980)$,
clearly visible in Fig.~\ref{fig.200524.1}, are generated dynamically from the
interactions between the pseudoscalars. The free parameter is the three-momentum cut-off
with natural size $\Lambda\simeq 1~$GeV used in the evaluation of the unitarity-loop functions
$g_{\Lambda i}(s)$ employed in this study.

It is also the case in some instances \cite{nd,jamin,guo.181123.1,guo.190126.1,guo.190126.3,alba.oller.glue}
that the ChPT expansion is complemented with the exchange of bare resonance fields, so that the tree-level amplitude
is crossing symmetric.
One typically improves the convergence properties of the chiral expansion by including bare resonance fields
because of the (partial) saturation of the chiral counterterms
by the resonance exchanges \cite{EP}. Then,
the matching process is undertaken up to ${\cal O}(\hbar p^4,\hbar^2)$, which
means to neglect any two-loop contribution and any one-loop contribution beyond ${\cal O}(p^4)$.
In this way, one could consider one-loop contributions involving higher orders because of the explicit inclusion of the
resonance fields.
The matching proceeds as in Eqs.~\eqref{200519.1} and \eqref{200519.2}, with the difference that now
the LO amplitudes 
include also the tree-level exchange of resonances  
and $T_4$ involves the one-loop contributions up to ${\cal O}(p^4)$. 
Then,
the Eq.~\eqref{200519.3} still holds and one has again Eq.~\eqref{200518.1} for $T_L$. 
This equation is in appearance analogous to the $N/D$-method form of Eq.~\eqref{A.4}.
Indeed, if we  identify $N_L$ with $V_L$ and $D_L$ with $I+V_L g(s)$, it
can be shown \cite{oller.review.old} that up to ${\cal O}(\hbar p^4,\hbar^2)$ the resulting functions satisfy
the $N/D$-method equations, cf.~Eq.~\eqref{eqs1}.

The perturbative solution of the $N/D$ equations with respect to the LHC contributions can also be
organized as an iterative solution in increasing number of insertions of $\Delta_\ell$. The first-iterated $N/D$ method
consists on taking only one power of $\Delta_\ell$ in the integrand of the DRs for $D_\ell(s)$ and $N_\ell(s)$.
The approximation is obtained by settling $D'_\ell(s)=1$ into the integrand for the DR of $N'_\ell(s)$, Eq.~\eqref{n2},
which is then denoted as $N'_{\ell;1st}$. Then,  
\begin{align}
\label{n2b}
{N}'_{\ell;1st}(s)=\sum_{m=0}^{n-\ell-1} \overline b_m s^m+\frac{(s-s_0)^{n-\ell}}{\pi}\int_{-\infty}^{s_{\rm Left}} ds' 
\frac{\Delta_\ell(s') }{p(s')^{2\ell} (s'-s_0)^{n-\ell} (s'-s)}~.
\end{align}                         
Since $\Delta_\ell(s)$ is known the DR integral could in principle be calculated. This is usually a tree-level amplitude
that can also be calculated in QFT, from which indeed $\Delta_\ell(s)$ is actually derived.
Therefore, we assume that in the first iterated $N/D$ method $N'_{\ell;1st}$ is also given.
As a result, the calculation of $D'_\ell(s)$ in this approximation, denoted by $D'_{\ell:1st}(s)$, just reduces
to perform the integration
\begin{align}
\label{d'b}
{D}'_\ell(s)=\sum_{m=0}^{n-1}\overline a _m s^m-\frac{(s-s_0)^n}{\pi}\int^\infty_{s_{\rm th}} ds' 
\frac{p(s')^{2L} \rho(s') {N}'_{\ell;1st}(s')}{(s'-s)(s'-s_0)^n}~.
\end{align}
The first-iterated $N/D$ method was used in Ref.~\cite{igi} to discuss $\pi\pi$ scattering within linear realizations
of chiral symmetry, taking into account the exchanges of a $\sigma$ and $\rho$ resonances.
More recently, it has been employed to study $\rho\rho$ scattering in Ref.~\cite{gulmez.rr}
by taking the pure gauge-boson part
of the non-linear chiral Lagrangian with hidden-local symmetry \cite{38gulmez,39gulmez}. Its generalization to
the $SU(3)$-related vector-vector scattering was undertaken in Ref.~\cite{gulmez.guo}.
These studies were motivated by the earlier ones in Refs.~\cite{rr.review}, with still an on-going
productive discussion in interpreting the results.

\subsection{FSI}
\label{sec.200517.3}

Let us consider the unitarity relation for a form factor, Eq.~\eqref{200220.5},
with the expression of the $T$ matrix in PWAs $T_L$ as given in Eq.~\eqref{200518.1}.
It then results that
\begin{align}
\label{181119.2}
F(s)&=\left(V_L^{-1}+g\right)^{-1}\left(V_L^{-1}+g+2i\rho(s)\theta \right)F^*~.
\end{align}
Since  $\Im g(s)=-\rho(s)$ it is clear that $g(s)+2i\rho(s)\theta(s)=g(s)^*$,
so that from Eq.~\eqref{181119.2} we have that along the RHC it is fulfilled that    
\begin{align}
\label{181119.3}
\left(V_L^{-1}+g\right)F&=\left(V_L^{-1}+g^* \right)F^*~.
\end{align}
The cancellation of $V_L$ from both sides leads to
\begin{align}
\label{181119.4}
\left[I+V_L(s) g(s)\right]F(s)&=\left[I+V_L(s) g(s)^* \right]F(s)^*~.
\end{align}
From this equation it is clear that the combination 
\begin{align}
\label{181119.5}
\left[I+V_L g(s)\right]F(s)
\end{align}
has no RHC \cite{basdevant.181119.1}. Then  $F(s)$ can be expressed as
\begin{align}
\label{181119.6}
F(s)&= \left[I+V_L(s) g(s)\right]^{-1}L(s)~,
\end{align}
with $L(s)$  a column vector of $n$ functions without RHC, being $n$ the number of PWAs.

An analogous relation can be obtained if we write $T_L(s)$ as in the $N/D$ method in coupled channels,
$T_L(s)=D_L^{-1}(s)N_L(s)$. Following the same steps as in Eqs.~\eqref{181119.2}--\eqref{181119.6},
taking into account that $\Im D(s)=- N(s)\rho$, one ends with the relations
\begin{align}
\label{181120.1b}
D(s)F(s)&=D(s)^*F(s)^*~,  \\
F(s)&=D(s)^{-1}L(s)~,\nn
\end{align}
and $L(s)$ is free of RHC. We can then write $F(s)$ as the product of 
two matrices, the inverse of $D_L(s)$, which only has RHC, and $L(s)$, which could have LHC. 
As a result,  Eq.~\eqref{181120.1b} is the generalization of the $N/D$ method to production processes.  

Coming back to Eq.~\eqref{181119.6}, let us remark that $I+V_L(s) g(s)$
could have the two types of cuts (since $V_L(s)$ in general has LHC).
For instance, for the case of the pion form factor if this is expressed as in Eq.~\eqref{181120.1b} 
then $L(s)$ has no LHC, while if expressed as in Eq.\eqref{181119.6}
it would typically have one, if $V_L(s)$ has it.
However, for the relevant case for phenomenological applications in which $V_L(s)$
is driven by the $s$-channel dynamics and it
does not comprise explicit LHC, then $L(s)$ has no either LHC.
In this case, the matrix $D_L(s)$ and $I+V_l(s) g$ can be identified.

This formalism has been employed by Ref.~\cite{oller.gama}
to study the $\gamma\gamma\to$meson-meson fusion reactions.   
Refs.~\cite{ozi,oller.D} used it to study the scalar form factor of the pion
(and of other pseudoscalar mesons) in connection with $J/\psi$ and $D$ decays,
and Ref.~\cite{palomar} analyzed the vector form factor of the pion.
This formalism was also very important to unveil the two-pole structure of
the $\Lambda(1405)$ in Ref.~\cite{plb}, because
in previous studies the $\pi\Sigma$ event distribution
for this resonance was always taken to be proportional to the
modulus squared of the  $\pi\Sigma\to\pi\Sigma$ $I=0$ $S$-wave.

\subsection{The exact $N/D$ method in NR scattering}
\label{sec.200517.4}

For non-relativistic scattering one can calculate for a given potential the exact discontinuity of a PWA along
the LHC. This has been a recent advance in $S$-matrix theory achieved by Ref.~\cite{oller.lhc.aop}, to which
we refer the reader for further details.
The key point was to extrapolate analytically  the LS equation to complex three-momenta for off-shell
scattering. The solution of the LS equation for half-off-shell scattering is an analytical function in the
off-shell three-momentum complex $q$ plane with vertical cuts which extend along the lines
$(\pm) p\pm i\lambda$, with $|\lambda|\geq \mu_0$. Here the $\pm$ symbols are unrelated,
$\mu_0$ is the lightest particle exchanged, and  
$p$ is the on-shell three-momentum (fixed by the energy $E$ of the process,
$E=p^2/2\mu$, with $\mu$ the reduced mass).
E.g. for $NN$ scattering the lightest particle exchange is the pion and $\mu_0=m_\pi$.
We denote in the following a PWA for half-off-shell scattering as $T_\ell(q,p)$, where $q$ is the off-shell
three-momentum and $p$ the on-shell one.

The discontinuity we are interested in, e.g. for its later application to the $N/D$ method, is 
\begin{align}
\label{200520.1}
\Delta_\ell(p^2)&=\frac{1}{2i}\left[T_\ell(p+i\ep,p+i\ep)-T(p-i\ep,p-i\ep)\right]=\Im T_\ell(p+i\ep,p+i\ep)~.
\end{align}
After some mathematical derivations that can be consulted in Ref.~\cite{oller.lhc.aop}, this discontinuity can be
obtained by  solving an ordinary linear IE.
This IE is written in terms of the discontinuity of the potential in momentum space $v_\ell(q,p)$. 
Its writing gets simplified by using $\hat{v}_\ell$ defined by
\begin{align}
\label{200520.2}
\hat{v}_\ell(q',q)&={q'}^{\ell+1}v_\ell(q,q)q^{\ell+1}~.
\end{align}
The discontinuity of the potential entering into the IE is
\begin{align}
\label{200520.3}
\Delta \hat{v}_\ell(\nu,\nu_1)&=
\Im \hat{v}_\ell(i\nu+\ep^-,i\nu_1+\ep)
-\Im \hat{v}_\ell(i\nu+\ep^+,i\nu_1+\ep)~,
\end{align}
with $\ep^-<\ep<\ep^+$ and $\ep^+\to 0$ at the end of the calculation. 
After this preamble, the sought IE is ($p=ik$, $k\geq \mu_0$ and  $n=2\ell+2$) 
\begin{align}
\label{200520.4}
f(\nu)&=\Delta \hat{v}_\ell(\nu,k)+\frac{\theta(p-2\mu_0-\nu)\mu}{2\pi^2}\int_{\mu_0+\nu}^{k-\mu_0}\frac{d\nu_1\nu_1^2}{k^2-\nu_1^2}
\left\{\frac{1}{(i\nu_1+0^+)^n}+\frac{1}{(i\nu_1-0^+)^n}\right\}
\Delta\hat{v}_\ell(\nu,\nu_1)f(\nu_1)~.
\end{align}
In terms of $f(\nu)$ the discontinuity $\Delta_\ell(p^2)$ is given by
\begin{align}
\label{200520.5}
\Delta_\ell(p^2)&=(-1)^\ell \frac{f(-k)}{2k^{2\ell+2}}~.
\end{align}
Thus, we need to solve the IE  for $\nu\in[-k+\mu_0,k-\mu_0]$, and the range of the integration in
the IE for $f(\nu)$ is finite for a given $p$, contrary to the LS equation. 
This IE can be solved without ambiguity because $\Delta\hat{v}(\nu,\nu_1)$ can be
determined for a given potential   and with it $f(\nu)$ by solving Eq.~\eqref{200520.4}.

For a general potential it is convenient to employ its spectral decomposition,
\begin{align}
v(\vq,\vp)&=\int_{\mu_0}^\infty d\bar{\mu}^2\frac{\eta(\bar \mu^2)}{(\vq-\vp)^2+\bar{\mu}^2}+\ldots
\end{align}
where $\eta(\bmu^2)$ is the spectral function, and the ellipsis indicates possible subtractions
that due to its polynomial nature do not give contribution to the discontinuity of the potential.  
In terms of the spectral decomposition we can write that 
\begin{align}
\label{200520.7}
\Delta \hat{v}_\ell(\nu,\nu_1)&=-\frac{2}{\pi}\int_{\mu_0}^\infty d\bar{\mu}^2 \eta(\bar \mu^2)\rho(\nu^2,\nu_1^2;\bar{\mu}^2)
\theta(\nu_1-\nu-\bar{\mu})~.
\end{align}
The function $\rho(\nu^2,\nu^2;\bar{\mu}^2)$ is a polynomial in its argument and its fixed by the
partial-wave projection involved in the case of interest. 
For brevity in the presentation offered here  we have just referred to the uncoupled case, but the formalism
can also be generalized easily to evaluate the LHC discontinuity for coupled PWAs \cite{oller.lhc.aop}.

A potential is said to be singular if for $r\to 0$ it diverges stronger than $1/r^2$ or as
$\alpha/r^2$ for $\alpha+\ell(\ell+1)<-1/4$. In the opposite case the potential is said to be
regular \cite{oller.lhc.aop}. In the ChPT calculation of nuclear potentials the increase in the order of the calculation
implies typically an increase in the degree of divergence of the potential for $r\to 0$, because
off-shell momentum factors give rise to spatial derivatives. This fact is the main reason why
the original Weinberg's program for solving nuclear properties once the chiral potentials are
calculated order by order has not been taken to full completion.

The resulting $\Delta_\ell(p^2)$ obtained by solving the master Eq.~\eqref{200520.4} has a different qualitative
behavior depending on whether the potential is regular, attractive singular or repulsive singular.
General arguments, based on the scaling properties of the  function $\rho(\nu^2,\nu_1^2;\bmu^2)$, 
were given in Ref.~\cite{oller.lhc.aop} to explain such differences in the behavior of $\Delta_\ell(p^2)$. 
Explicit examples were also worked out in Ref.~\cite{oller.lhc.aop} corresponding to actual PWAs in
$NN$ scattering, with the chiral potential calculated at different chiral orders, from LO up to NNLO.
The function  $\rho(\nu^2,\nu_1^2;\bmu^2)$ is a polynomial in $\nu$ and $\nu_1$ of degree $m$.
Then, the argument of Ref.~\cite{oller.lhc.aop} follows by considering  a re-scaling
by a parameter $\tau$ of the variables $k$, $\nu$ and $\nu_1$ in the limit  $k\gg\mu_0$.
It follows from Eq.~\eqref{200520.4} that the $n_{th}$ iterated solution for $f(\nu)$
is subject to a  re-scaling by
\begin{align}
\label{200520.8}
\tau^{(n+1)m-(2\ell+1)n}=\tau^{(m-2\ell-1)n+m}~.
\end{align}
The point is whether $m-2\ell-1$ is smaller or larger than zero. In the former case we have the
behavior corresponding to a regular potential, so that each extra iteration implies at least
an extra factor of $1/k$ and for $k\to\infty$ the discontinuity $\Delta_\ell(-k^2)$ tends to its Born
approximation. However, when $m-2\ell-1>0$ each iteration increases the power of $k$ in the asymptotic behavior
of $\Delta_\ell(-k^2)$, becoming more and more divergent as $n$ increases.
This is the situation for a singular potential. 

For the regular potentials $\Delta_\ell(p^2)$ tends to its Born term contribution which vanishes at least as
$1/p^2$ for $p^2=-k^2$ and $k^2\to \infty$.
For such type of $\Delta_\ell(p^2)$ it was shown in Ref.~\cite{guo.rios.NN}
that any $N/D$ IE, irrespectively of the number of subtractions taken, has solution.
However, for singular potentials the resulting $|\Delta_\ell(p^2)|$ grows faster than any polynomial in
the same limit.  This is clearly shown in Ref.~\cite{oller.lhc.aop} by log-log plots in which the
slop of $|\Delta(-k^2)|$ continuously grows with increasing $k^2$. As a dramatic consequence of this result is that
it is not possible to write down a DR representation for a NR PWA if the potential is singular. However,
it is still possible to use the $N/D$ method  because what matters for the $N/D$ IEs is the product
$\Delta_\ell(-k^2)D_\ell(-k^2)$. The denominator function is known to behave asymptotically
as $s^{-\delta(\infty)/\pi}$, cf. Eq.~\eqref{181122.3}, and $\delta(\infty)=N\pi$, 
with $N$ the number of bound states,  because of the Levinson theorem.
It turns out that the number of such stats is infinite for attractive singular potentials \cite{singular.rev}
and, in this case, $D_\ell(-k^2)$ vanishes also faster than any power law.

The exact $N/D$ method is defined in Ref.~\cite{oller.lhc.aop} as the $N/D$ method
but using $\Delta_\ell(p^2)$ stemming from
Eqs.~\eqref{200520.4} and \eqref{200520.5}, which is the exact LHC discontinuity of the full PWA
for a given potential.
In this way, we showed in Ref.~\cite{oller.lhc.aop} that one reproduces exactly the
LS-equation solutions for regular potentials. This is also true for the singular potentials
when the potential is used in the whole range of integration
in the LS equation, that is, for $q\in[0,\infty]$ (the cut-off is sent to infinity).
For the singular-potential case we refer to the standard kind 
of solutions, so that for a repulsive singular potential the solution has no free
parameters and is determined,
while for the attractive singular case the solution involves one free parameter that could be fixed e.g.
by imposing a given value for the scattering length \cite{singular.rev,singular.case,singular.arr}.
Several potentials were studied in Ref.~\cite{oller.lhc.aop}, both for uncoupled and coupled PWAs.
Within the latter group the $^3S_1$--$^3D_1$ coupled PWAs were studied and the $^3S_1$ scattering length was
taken as input. Needless to say,  in all cases the LS equation with infinite cutoff and the $N/D$ method agree
perfectly in our numerical study.

The fact of having none or only one free parameter is a very constrained situation in practical applications,
and it is the reason why it has not been possible to achieve yet a good
agreement with data in  $NN$ scattering in terms of regulator-independent solutions (i.e. in
which the three-momentum cut-off is taken to infinity). Notice that the number of free parameters in the solution
of the LS equation for singular potentials is then not linked with the chiral order
in the calculation of the chiral potential. 
However, in terms of the $N/D$ method one can in principle add an arbitrary number of subtractions,
which allows one to look for  extra solutions. 
We have already discussed this point in connection with the ambiguity associated with the CDD poles
in Sec.~\ref{sec.200517.2}.
This possibility was explored in detail in Ref.~\cite{oller.lhc.plb} for the $^1S_0$ $NN$ PWA.
The NLO and NNLO ChPT potentials for this PWA are actually attractive and singular. The standard solutions of
the LS equation were reproduced, and a detailed numerical analysis was performed in order to show the 
agreement between the LS equation and the exact $N/D$ method.
But we also showed in this reference that one can generate new solutions that cannot be achieved
by the LS equation when the three-momentum cut-off is taken to infinity 
with contact interactions included in the potential to aim renormalization (in the form of
polynomial counterterms in its momentum expression).
In this way, a new solution was discussed that can reproduce the $^1S_0$ 
scattering length, effective range and shape parameter $v_2$. 
For this solution the DRs for $N_\ell(s)$ and $D_\ell(s)$ converge separately. 
It is also interesting to indicate that a solution within the exact $N/D$ method for this PWA
fixing only two parameters, the scattering length and the effective range were taken, could not be found. 
Last but not least, a very attractive feature of the exact $N/D$ method is that it allows to evaluate the PWAs in 
the whole complex $p^2$ plane. Then, it is very convenient to look for resonance and (anti)bound states.
In the case of the $^1S_0$ PWA there is an antibound state which is found at $p=-i0.066$~MeV both at NLO and NNLO
when all the first three ERE parameters are reproduced.

\section{Conclusions}
\label{sec:conclu}
\def\theequation{\arabic{section}.\arabic{equation}}
\setcounter{equation}{0}

We have elaborated on several unitarization methods of perturbative calculations
in Chiral Perturbation Theory (ChPT) that can be employed to study scattering and
the re-scattering corrections to an external probe. Special attention has been
given to the $N/D$ method both for scattering and for implementing  the final-state interactions (FSI).
The unitarization methods, since the earlier papers on current algebra techniques,
have been able to extend to much larger energies the expected region of utility of
ChPT calculations.
This has been accomplished thanks to the extra energy and momentum dependence
generated by using a non-perturbative
theoretical framework which satisfies key properties of $S$-matrix theory,
which stem from two-body unitarity and analyticity.
Some of the most striking and important applications of the unitarization methods of input perturbative
calculations have occurred in the field of spectroscopy.
 In this way,  
it has been possible to study resonances and bound states, and even predict
some of them, while unexpected properties have been unveiled too, as e.g. the two-pole nature
of some resonances \cite{meissner.two}, as first shown for the $\Lambda(1405)$ in Ref.~\cite{plb}.

Along this review we have paid attention to establish links between  different unitarization methods.
Thereby, by starting with the (generalized) relativistic effective-range expansion (ERE)
we have connected it with the $K$-matrix approach and then obtained from the former
the Inverse Amplitude Method (IAM) unitarization formula. 
The IAM  has been also connected with the Padd\'e approximation.
In the last part of the manuscript we have introduced and discussed the $N/D$ method.
A link between the $N/D$ and the IAM can also be established by employing
the solution to the $N/D$ method based on treating perturbatively the left-hand cut discontinuity.
This allows one to derive the IAM as a particular case of this method too. 
The associated methods to take care of the FSI
corresponding to the unitarization techniques of scattering have been introduced as well. In addition,
we have discussed the (Muskhelishvili-)Omn\'es solution and the Khuri-Treiman approach.  

An advantage of the unitarization technique based on the $N/D$ method is that it
can be applied to deliver the 
unitarized partial-wave amplitudes (PWAs)
even if only the leading-order scattering amplitudes are employed.
A subtraction constant is then required, but it could be estimated making use of naturalness arguments.
In this way, one can study important resonances in hadron physics in a very constrained manner,
essentially without any free parameter. Good examples are 
the $f_0(500)$, $f_0(980)$, $a_0(980)$, and $\kappa(800)$ in the scalar light mesonic sector, the
$\Lambda(1405)$ in the  strangeness $-1$ $S$-wave meson-baryon scattering, etc. 
Of course, one could also use as input higher-order scattering amplitudes  provided by the effective field theory
of interest and perform a higher-order analysis in the input taken.

Regarding the $N/D$ method, we would like to stress that thanks to recent advances (in which the author has been involved),
it can be considered for non-relativistic scattering as an alternative formulation of scattering theory.
The qualitative leap forward has been the derivation of the exact discontinuity of a PWA along the
left-hand cut, which can then be employed to solve the $N/D$ integral equations. In this way, one can solve
standard regular potentials and reproduce the solutions obtained with the Lippmann-Schwinger equation
for PWAs.
But it also allows to obtain extra solutions for singular potentials without dependence on cutoff, which can be sent to infinity.
This method has the advantage that, in terms of the solution found, it is 
straightforward to evaluate the on-shell scattering amplitudes in the complex energy plane too.
In this way, e.g. one could look  for poles and their residues
(which give the resonance or bound-state couplings).
This is a very promising and exciting field of current research,
and first applications are being explored for $NN$ scattering. 

\acknowledgments{This work has been supported in part by the MEC (Spain) and FEDER (EU) Grants FPA2016-77313-P and
PID2019-106080GB-C22.}

\reftitle{References}

\end{document}